# Visualization of Protein 3D Structures in Reduced Representation with Simultaneous Display of Intra- and Inter-Molecular Interactions*

**Vrunda Sheth[1] and Vicente M. Reyes, Ph.D.[2], [3]**

(*, M. S. Thesis; [1], M. S. student; [2], thesis advisor)
[3], E-mail: **vmrsbi.RIT.biology@gmail.com**

**Vrunda Sheth**

Dept. of Biological Sciences, School of Life Sciences
Rochester Institute of Technology
One Lomb Memorial Drive, Rochester, NY 14623

October 2009

# Visualization of protein 3D structures in reduced representation with simultaneous display of intra and inter-molecular interactions

Vrunda Sheth

# *Visualization of Protein 3D Structures in Reduced Representation with Simultaneous Display of Intra and Inter-molecular Interactions*

Vrunda Sheth

Approved: \_\_\_\_\_\_\_\_\_\_\_\_\_\_\_\_\_\_\_\_\_\_\_\_\_\_\_\_\_\_\_\_\_\_
**Vicente Reyes, Ph.D.**
*Thesis Advisor*

\_\_\_\_\_\_\_\_\_\_\_\_\_\_\_\_\_\_\_\_\_\_\_\_\_\_\_\_\_\_\_\_\_\_
**Gary Skuse, Ph.D**.
*Committee Member*

\_\_\_\_\_\_\_\_\_\_\_\_\_\_\_\_\_\_\_\_\_\_\_\_\_\_\_\_\_\_\_\_\_\_
**Paul Craig**, **Ph.D.**
*Committee Member*

*This thesis is dedicated to my beloved family, to my parents for their never ending encouragement and confidence in me and to my sisters for their motivation and guidance.*

**DISSERTATION AUTHOR PERMISSION STATEMENT**

**TITLE OF THESIS:** Visualization of protein 3D structure in a reduced representation with simultaneous display of intra and inter-molecular hydrogen bonds and van der Waals interactions.

Author: Vrunda Sheth

Degree: Masters

Program: Bioinformatics

College: College of Science, Rochester Institute of Technology

I, Vrunda Sheth, understand that I must submit a print copy of my thesis or dissertation to the RIT archives, per current RIT guidelines for the completion of my degree. I hereby grant to the Rochester Institute of Technology and its agents the non-exclusive license to archive and make accessible my thesis or dissertation in whole or in part in all forms of media in perpetuity. I retain all other ownership rights to the copyright of the thesis dissertation. I also retain the right to use in future works (such as articles or books) all or part of this thesis or dissertation.

_______________________________
Vrunda Sheth

Date

**DISSERTATION PRINT REPRODUCTION PERMISSION STATEMENT**

**TITLE OF THESIS:** Visualization of protein 3D structure in a reduced representation with simultaneous display of intra and inter-molecular hydrogen bonds and van der Waals interactions.

Author: Vrunda Sheth

Degree: Masters

Program: Bioinformatics

College: College of Science, Rochester Institute of Technology

I, Vrunda Sheth, hereby grant permission to the Wallace Memorial Library, of Rochester Institute of Technology, to reproduce my print thesis in whole or in part. Any reproduction will not be for commercial use or profit.

$\overline{\phantom{x}}$

Vrunda Sheth

Date:

**Visualization of protein 3D structure in a reduced representation with simultaneous display of intra- and inter-molecular hydrogen bonds and van der Waals interactions.**

Vrunda Sheth

October, 2009

Approved:

_______________________________
Dr. Vicente Reyes
Thesis Advisor,

_______________________________
Dr. Gary Skuse
Director, Bioinformatics

_______________________________
Dr. Paul Craig
Committee Member

# ACKNOWLEDGEMENTS

I feel it is a unique privilege, combined with immense happiness, to acknowledge the contributions and support of all the wonderful people who have been responsible for the completion of my masters degree. The two years of graduate study at RIT has taught me that creative instinct, excellent fellowship and perceptiveness are the very essence of science. They not only impart knowledge but also place emphasis on the overall development of an individual. I am extremely appreciative of RIT, especially the Department of Biological Sciences (Bioinformatics Options) in this regard. I owe it to my mentors at RIT to what I am today.

I would like to express my deepest gratitude to my thesis advisor, Dr Vicente Reyes who continually encouraged and guided me during the course of my thesis work. I would also like to express my appreciation to my committee members, Dr. Gary Skuse and Dr. Paul Craig for their valuable guidance, timely help and support. I would also like to thank Dr. Gurcharan Khanna, Ryan Lewis, Thomas Batzold and all other members of RIT Research Collaboratory for allowing me to use their advanced computation resources and for their timely help and guidance. I would also like to thank Kyle Dewey for his continued help during the course of my thesis. A special thank to Nicoletta Bruno Collins for all the academic formalities that needed to be done.

I would like to thank all professors, mentors, family, friends and well wishers who have helped me scale heights and achieve this prestigious degree at RIT. Finally, I would like to thank RIT for giving me an opportunity to be a part of its great family.

# LIST OF ABBREVIATIONS

| PDB | Protein Data Bank |
|---|---|
| ID | Identification |
| HTML | HyperText Markup Language |
| PHP | Hypertext Preprocessor |
| MATLAB | Mathematics Laboratory |
| AAR | All Atom Representation |
| DCRR | Double Centroid Reduced Representation |
| H-bonds | Hydrogen Bonds |
| VDW | van der Waals Interaction |
| CPK | Corey Pauling Koulton |
| GUI | Graphical User Interface |
| SCOP | Structural Classification of Proteins |
| 3D | Three Dimensional |
| C-α trace | Backbone alpha Carbon atoms |
| traM | Transcriptional Repressor |
| Fe | Iron |
| CA | Calcium |

# ABSTRACT

Protein structure representation is an important tool in structural biology. There exists different methods of representing the protein 3D structures and different biologists favor different methods based on the information they require.

Currently there is no available method of protein 3D structure representation which captures enough chemical information from the protein sequence and clearly shows the intra-molecular and the inter-molecular H-bonds and VDW interactions at the same time. This project aims to reduce the 3D structure of a protein and display the reduced representation along with inter-molecular and the intra-molecular H-bonds and van der Waals interactions. A reduced protein representation has a significantly lower "atomicity" (i.e., number of the coordinates) than one which is in all-atom representation. In this work, we transform the protein structure from 'all-atom representation' (AAR) to 'double-centroid reduced representation' (DCRR), which contains amino acid backbone (N, Cα, C', O) and side chain (Cβ and beyond) centroid coordinates instead of atomic coordinates.

Another aim of this project is to develop a visualization interface for the reduced representation. This interface is implemented in MATLAB and displays the protein in DCRR along with its inter-molecular, as well as intra-molecular, interaction. Visually, DCRR is easier to comprehend than AAR. We also developed a Web Server called the Protein DCRR Web Server wherein users can enter the PDB id or upload a modeled protein and get the DCRR of that protein. The back end to the Web Server is a database which has the reduced representation for all the x-ray crystallographic structure in the PDB.

# Contents

**LIST OF FIGURES**

**LIST OF TABLES**

# Chapter 1
Introduction

**1 Background**

*1.1 Methods of structure representation.*

Protein structure representation is an important area in structural biology. The function of a protein is determined by its 3D structure (Kim et al., 2006). Hence, ways to visualize protein and compare protein structures eventually paves the way to protein function annotation. There are several methods of protein structure representation. Described below are a few common methods:

1.1.1 α-carbon backbone

This representation of the protein shows only the position of the backbone α-carbon atom. It does not show other backbone atoms and side-chain atoms. This structure gives minimal information about protein interactions, and its chemical properties as embodied by the side-chain of the amino acid residues. However, it clearly shows the secondary structure of the protein (Sayle & Milner, 2000).

1.1.2 all-atom models

The wireframe and the ball and stick model are examples of all-atom representation model (AAR). The wireframe model represents the bond between the atoms. In this representation all the atoms (usually the non-hydrogen atoms) of a protein are displayed. However, it is difficult to identify the secondary structure. It shows a crowded display of the protein showing too much information simultaneously. The ball and stick model is similar to the wireframe, and it uses sticks for representing the bonds and small spheres for representing the atoms. Since all the atoms are displayed, this type of structure is very good for checking local details. However for larger structures this type of information is overwhelming and it is difficult to retrieve specific information. Moreover, this representation is not very good for viewing the secondary structure or protein fold information (Sayle & Milner, 2000)

### 1.1.3 Ribbon

This form of representation depicts α –helices as either ribbons or cylinders and β-sheets as arrows pointing in the N ⟶ C direction. This type of representation provides a very good view of the secondary structures in the protein, and connections between the secondary structures and is a good representation for protein structural comparison. It represents only the protein backbone and no side chain information is represented. Moreover, the ribbon representation does not show the H-bonds and the VDW interactions (Sayle & Milner, 2000).

### 1.1.4 Space filling models

This model represents all atoms as spheres with diameters proportional to the atomic radius of the element. It is also called the CPK model. Each atom is colored differently, i.e. hydrogen is white, nitrogen is blue, oxygen is red and sulfur is yellow. This model gives an idea of the overall representation of the protein. However, it is not useful for viewing the secondary structure or the active sites (Sayle & Milner, 2000).

## *1.2 Visualization tools*

There are also several visualization tools which allow protein to be visualized in one or more of the above representations, the most common among them being Rasmol (Sayle, 2000), Pymol (DeLano, 2002), Deep view (Guex, 1999), SWISS-MODEL (Torsten et al., 2003), KING (Chen et al.,2009). Most of these visualization tools have both command line and GUI interface, allowing ease of use. Some of these visualization tools can also be customized with user plug-in scripts. All of these visualization tools read PDB structure files and produce an output image of the protein structure. These images can be rotated, zoomed in and out, and colored differently according to user preference. Most of these visualization tools are interactive, allowing users to pick points to calculate distance between them or calculate the dihedral angle. Hence the visualization tools provide a depth of explorative information about the protein structure.

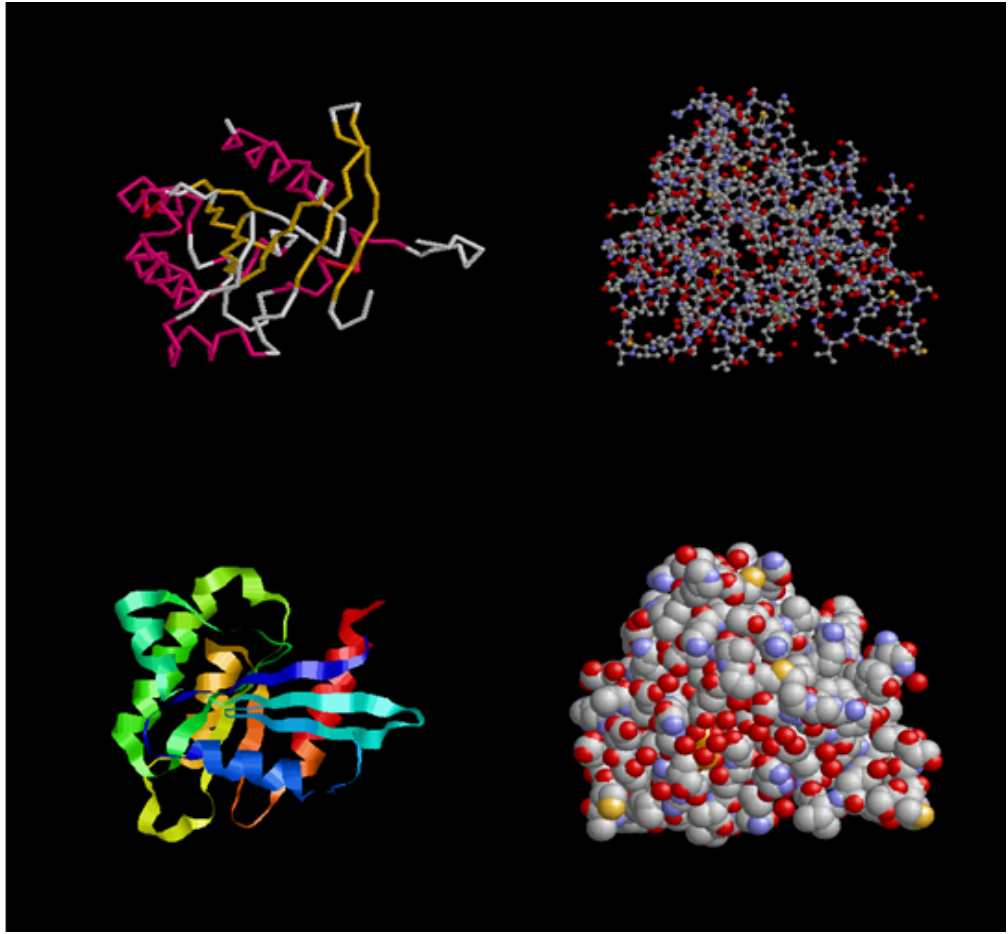

*Figure1: These are images of protein 2g3- human GTpase generated using Rasmol in different protein structure representation model. Above left, α-carbon backbone; above right, all-atom model; below left, ribbon model; below right, space filled model.*

*1.3 PDB*

The PDB is the central repository of all the protein 3D structures. PDB started in early 1970's with about seven structures. Today there are about 59,330 structures in the PDB, of which about 50,000 were solved using x-ray crystallographic methods (Berman et al., 2002)

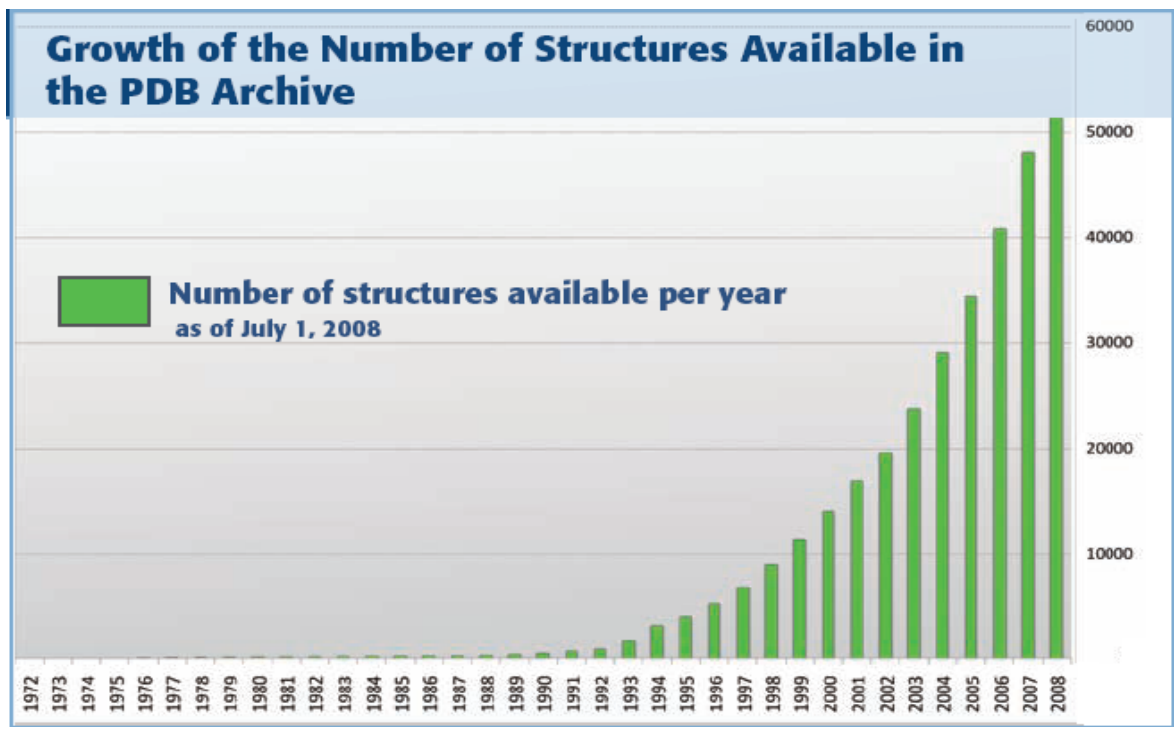

*Figure 2: Growth of PDB (PDB Annual Report, 2008)*
*<http://www.rcsb.org/pdb/general_information/news_publications/annual_reports/annual_report_year_2008.pdf>.*

The PDB is maintained by RCSB, which curates and annotates the PDB structures (Berman et al.,2002). The PDB provides structural information in various formats such as the PDB and mmCIF. The PDB also provides information about the sequence, sequence similarity, and biological and chemical properties of the protein along with links to various visualization tools like JMOL. The PDB is growing in size as new structures get deposited in it every week. (http://www.rcsb.org/pdb/home/home.do)

## 2. Motivation

The current protein structure representation models have their own set of pros and cons. The all-atom models like the wireframe and ball and stick model represent all the chemical information of the protein however the display is overwhelming with too much information presented at the same time. On the other hand, the surface representation models like the CPK space filling model provide an excellent view of the surface properties of the protein and locate the shape of complementary sites of protein-protein interactions. However the space representations fail to illustrate the secondary structures, the active site and the interactions. Finally the ribbon model provides a very good view of the secondary structures and the connection between them but do not show the intra and inter-molecular interactions.

We aimed to develop a protein 3D structure model that captures just enough chemical information to clearly display the secondary structure of the protein as well as the inter and the intra-molecular interactions. This protein representation model provides a balance between too much information and too little information. Our motivation led to the development of a model called the Double Centroid Reduced Representation or the DCRR model.

We used a few representative structures from the PDB for model development, testing and development of the visualization interface. We then converted all protein structures currently deposited (ca. Aug. 2009) in the PDB from AAR to DCRR.

# Chapter 2

Reduced representation of proteins

We have implemented a new method of protein structure representation called the double centroid reduced representation (DCRR) wherein each amino acid is represented by two data points, namely the centroid of its backbone atom and the centroid of its side-chain. The backbone atoms are N, Cα, C' and O while the rest like Cβ, Cγ, Cδ, and Cε are the side-chain atoms. This representation captures the chemical information in the protein by specifying the side-chain and backbone as centroids without it being too visually complex. In the DCRR model we also show the intra and the inter-molecular H-bonds and van der Waals interactions (VDW). DCRR also clearly depicts the ligands, ions and the bound ordered water molecules.

DCRR strikes a balance between AAR which displays too much chemical information precluding clear display, and the ribbon model which excludes too much chemical information, disallowing display of side chains residues.

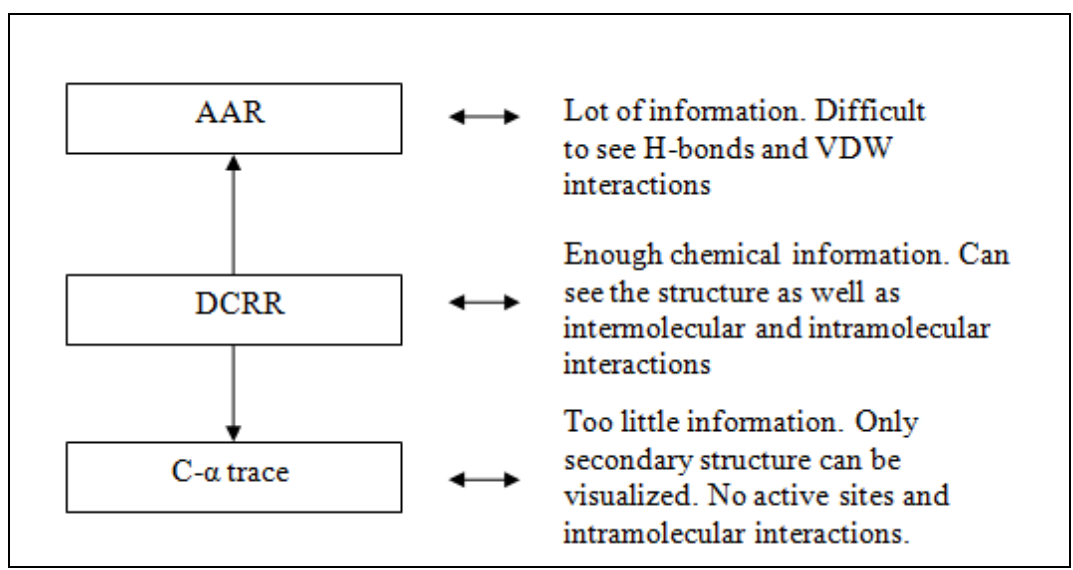

*Figure 3: Position of DCRR with respect to AAR and C-α trace.*

Figure 4 depicts the prototype we had before we started developing our model. We hoped to create a model which clearly depicts the secondary structure of the protein as well as all physical interactions (hydrogen bonds and van der Waals interactions).

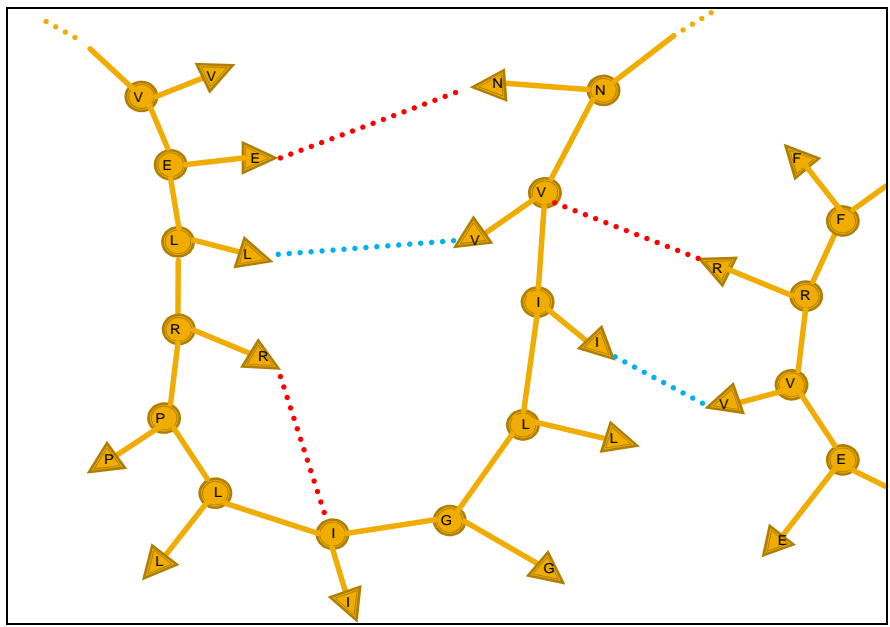

*Figure 4: DCRR of a hypothetical protein*

We also developed a graphical interface to display the reduced representation of protein. This interface was implemented in MATLAB and displays not only the reduced structure but also H-bonds and van der Waals interaction among the amino acid residues in the proteins as well as those between proteins and the ligands or the bound ordered water molecules. To the best of our knowledge, this is the first ever visualization interface for any reduced protein representation.

We developed a database complementary to the PDB. The database is housed on our bioinformatics cluster and it has the DCRR of all those structures in PDB which have been determined using X-ray crystallographic method.

Finally we have also developed a Web Server wherein users can enter the PDB id or upload a modeled protein and get the reduced representation for that protein. The Web Server is a PHP interface which queries the database using the PDB id provided by the user.

Chapter 3

Implementation

**1 Preprocessing steps**

To convert a protein into a reduced representation, it has to go through a preprocessing pipeline. This pipeline has several scripts which take the protein atom file from the PDB and converts it into a DCRR file. It has a series of scripts to calculate the intra and the inter-molecular H-bonds and van der Waals interactions.

*1.1 Transformation of AAR to DCRR*

This step involves running a series of scripts which first separates the backbone atoms from the side-chain atoms and then computes the 'geometric' or 'positional centroid' of the backbone and the side-chain atoms. This is simply the arithmetic mean of the x, y and z coordinates of the backbone and the side chain atoms of each amino acid (i.e., the different atoms are not weighed according to their atomic weights). For example consider a backbone of any amino acid:

| Atom | X | Y | Z |
|---|---|---|---|
| N | 10 | 12 | 18 |
| $C_\alpha$ | 12 | 17 | 23 |
| C' | 14 | 43 | 34 |
| O | 16 | 32 | 10 |

*Table 1: Backbone atoms*

The centroid of the backbone of this amino acid has x, y and z coordinates of 13, 26 and 21.25, respectively. In this way for every amino acid the backbone and the side-chain centroid is calculated.

## 1.2 Nearest –neighbor calculation

This step determines the neighboring atoms of every atom in a protein. It does this by sequestering all atoms that lie within a specified radius from a given atom. Typically we use radius=5.0 Å. For each atom in a protein all the atoms within 5.0 Å are considered to be its neighbors. Based on the distance and the identity of the atoms we then proceed to calculate the H-bonds and the van der Waals interactions.

## 1.3 Calculating the H-bond and the Vander Waals interactions.

One of the most important preprocessing steps is the calculation of the H-bonds and the van der Waals interactions. These were calculated using information about the atomic identity of nearest neighbors and their distance from each other (2.80 Å for hydrogen bonds and 3.38 Å for -CH-HC- van der Waals interaction). Separate scripts are used to calculate the intra-molecular interactions which are bonds within a protein and the inter-molecular bonds between a protein and a ligand or protein and bound ordered water molecules.

## 1.4 Optimization of H-bonds and van der Waals interactions

In these optimization steps, we determined the appropriate number of H-bonds and van der Waals interactions to show in the display that is not too many to crowd the display and not too few to miss the important ones.

### 1.4.1 Optimization of H-bonds

In H-bond distance optimization, we used the standard H-bond length of 2.80 Å as obtained from the literature (Jeffery, 1997). In order to accurately capture all the H-bonds we have designed a window size that defines the upper and lower limits around this H-bond ideal length. The upper and lower bounds of the window were varied (increased and decreased) until an optimal number of H-bonds were obtained. After a series of simulations we finalized the optimal upper and lower limit for H-bond interaction to be 2.72Å and 3.22Å respectively. These will be our recommended limits for viewing the DCRR structure. Users will have an option to view wider limits or narrower limits depending on their analysis. The lower and upper bound for H-bonds are 2.66Å and 3.36Å for wide range and 2.75Å and 3.0Å for narrow limits respectively.

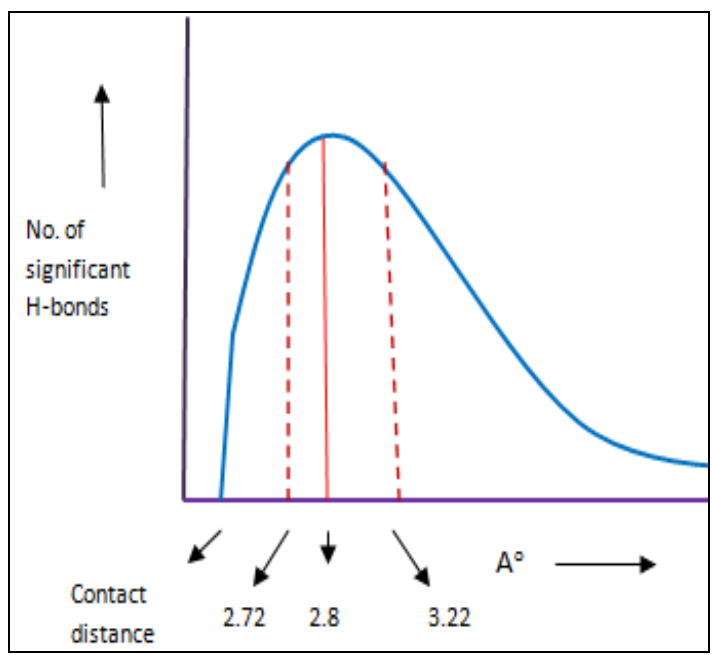

*Figure 5: The H-bond limits for the recommended range*

1.4.2 Optimization of van der Waals interactions

van der Waals interaction distance optimization is done in a similar way to the H-bond distance optimization. In the calculation of van der Waals interaction we are considering only the H--C-C--H van der Waals interaction. The standard van der Waals radius for H--C-C--H is 3.38 Å (Bondi, 1964). The window defining the upper and the lower limits was varied until an optimal number of van der Waals interactions for display were obtained. We finalized the recommended window size to be 3.20 Å as the lower limit and 3.85 Å to be the upper limit. Users can ask for a narrower or a wider limit based on their research needs. The lower and upper bound for van der Waals interactions are 3.10 Å and 3.95Å for wide range and 3.30 Å and 3.75 Å for narrow limits respectively.

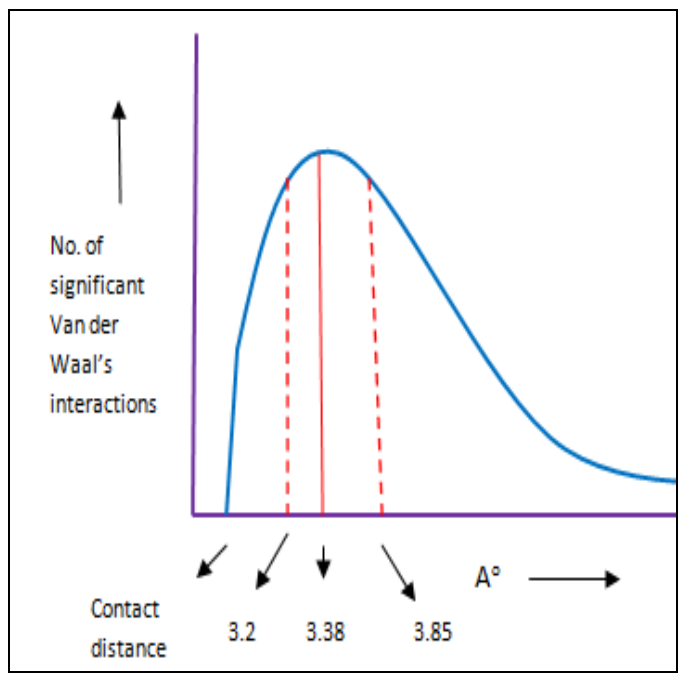

*Figure 6: The van der Waals limits for recommended range*

1.4.3 Optimization of bound ordered water molecules

The B-factor value in the PDB file was used for the optimization of the bound ordered water molecules. The B-factor is present for every atom in a protein PDB file. The B-factor is the number that gives an idea of the thermal vibration of the atom. The lower the B-factor value, the higher the probability that an ordered water molecule is bound at the indicated position. We have taken a B-factor cutoff of 18 to incorporate the most important water molecules that play a role in stabilizing the protein structure. This B-factor cutoff was determined using a series of empirical tests trying various values of B-factor until an optimal number of water molecules for display are obtained.

## 2. Visualization interface implementation

The visualization script for the DCRR of a protein is written in MATLAB. MATLAB was chosen not only because of its capability to create diverse kinds of expressive plots, but also its widespread availability and popularity. MATLAB figure files can be zoomed in, zoomed out, rotated and translated allowing a proper view of the protein 3D structure.

MATLAB can very easily be interfaced with other programming languages such as JAVA, PERL and python. MATLAB can call java libraries allowing easy programming. Hence MATLAB was a good choice for the development of a visualization interface.

MATLAB protein DCRR image clearly demarcates the secondary structures of the protein and also shows the H-bonds and the van der Waals interactions. The finished image has the following features:

1. The protein backbone centroids are connected in a black solid line.
2. The side chain centroids are represented as small spheres extending from the respective backbone centroids.
3. Each amino acid is labeled and identified with a single letter code.
4. Each amino acid is color coded based on its polarity.
5. H-bonds stabilizing the structure of the protein as well as those interacting with ligands and bound ordered water molecules, are shown in blue dashed lines.
6. van der Waals interactions in the protein and between proteins and ligands are shown as red dashed lines.
7. Ligands are shown in bright green triangles.

8. Bound ordered water molecules are shown in blue squares.

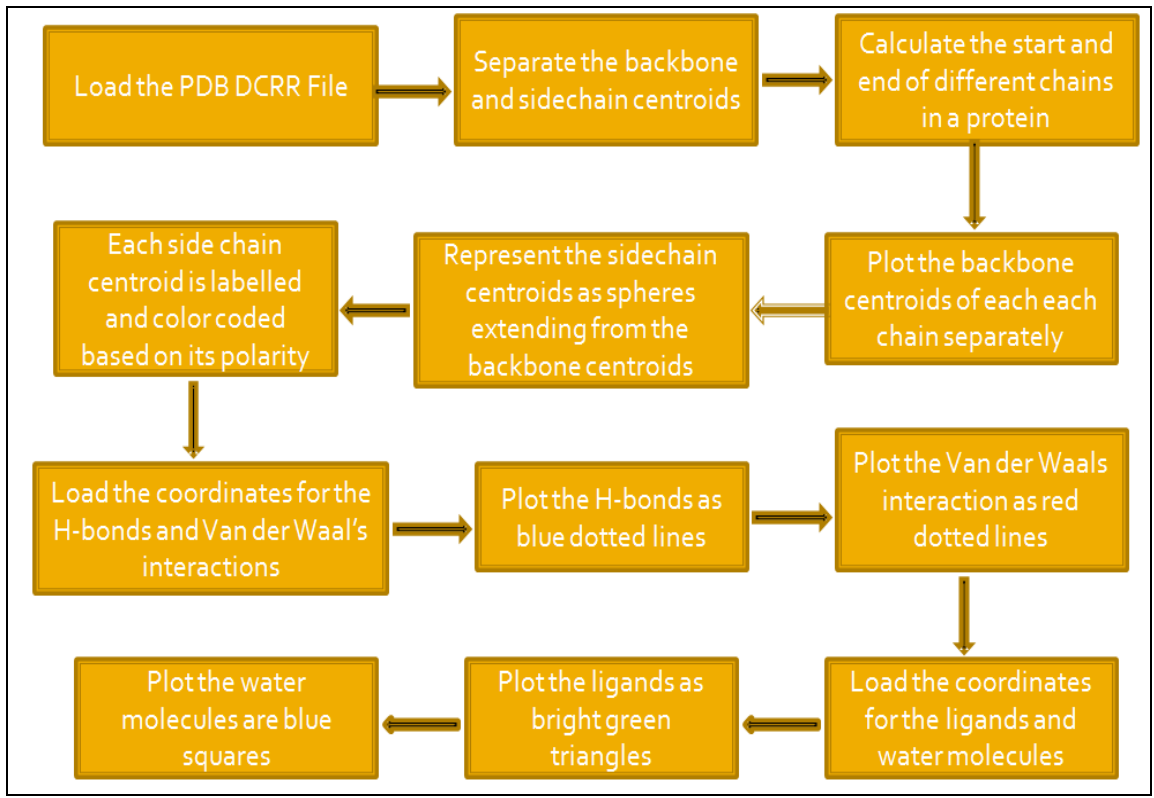

*Figure 7: Flow diagram representing steps in MATLAB*

Figure 8 shows a Venn diagram which was used as a reference to design the color-coding scheme for the amino acids.

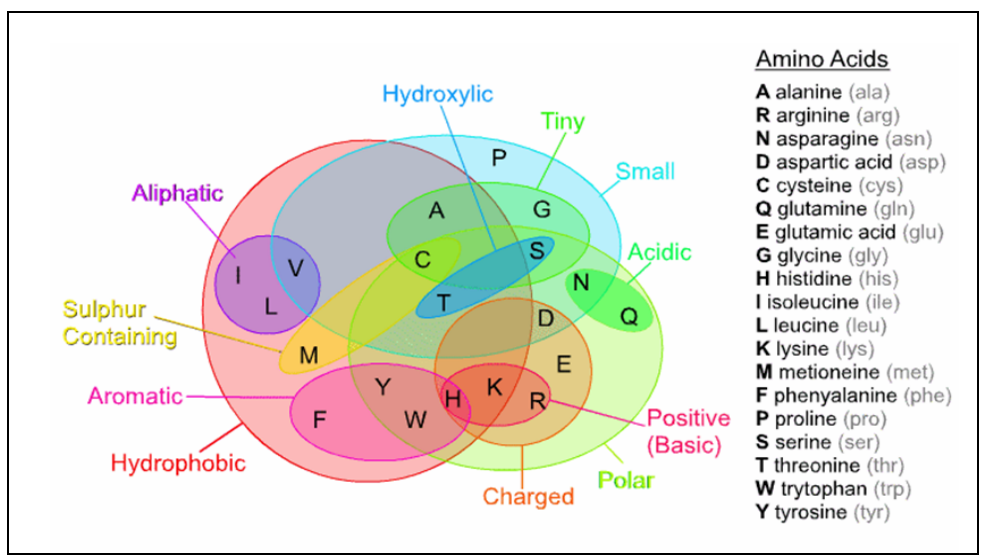

*Figure 8: Color coding scheme (Gelder & Vriend, 2008)*
*< http://swift.cmbi.kun.nl/teach/ALIGN/Align_8.html>.*

## 3. Web Server Implementation

We also developed a web server called the Protein DCRR Web Server. The Web programming language used was PHP and Ajax. This Web Server interfaces a MySQL database at the backend. In this Web Server, the user can enter the PDB id of any protein solved using x-ray crystallography. Users will be sent a link to a zipped folder. The folder contains the DCRR coordinate file, the coordinates of the H-bonds and van der Waals interaction, the coordinates of the ligands and the bound ordered water molecules and the DCRR image of the protein as a MATLAB figure file.

The figure file can be opened using MATLAB and the protein image can be rotated, translated, zoomed in and zoomed out.

Apart from entering a PDB id, the Web Server also allows users to upload a modeled protein whose structure has not yet been deposited in the PDB. The file for the modeled protein should be in the strict PDB format. This file will be validated by the server and then submitted to the DCRR pipeline. Users will get a link pointing to the result. The result will also be emailed to the users.

At the backend of the Web Server is a MySQL database which contains links to the reduced representation of all the X-ray crystallographic structures in the PDB. This database was created to reduce the redundancy of the job. All the PDB structures have been converted to a reduced representation and stored in the database. When a user wants to view the reduced structure the Web Server points the user to the folder containing the compressed file available for download. The Web Server and the database are housed on the Bioinformatics Server at RIT. The link to the webpage is http://tortellini.bioinformatics.rit.edu/vns4483/dcrr.php

*Figure 9: Home page of protein DCRR Web Server*

Several hints have been added to the web page to allow proper input. (*Ask the CSS guy*). To prevent faulty or dangerous file uploads both client and server side validation has been set up.

Client-side validation makes sure that either a PDB id of a protein or a file for the modeled structure of a protein is uploaded. It prompts the user in case both the PDB id as well as a file is uploaded or if neither field is filled. It checks to make sure that the PDB id is a 4 letter string. The Web Page also has several alert pop-ups to warn users of wrong input or processing time. If the user enters a protein id the result page is displayed immediately. However if a file is uploaded, the processing time can go from anywhere between 2-50 minutes depending on the size of the protein.

The PHP post method is used to send the data in an encrypted format to the database and query the database to retrieve user information or start a script to convert the structure to DCRR.

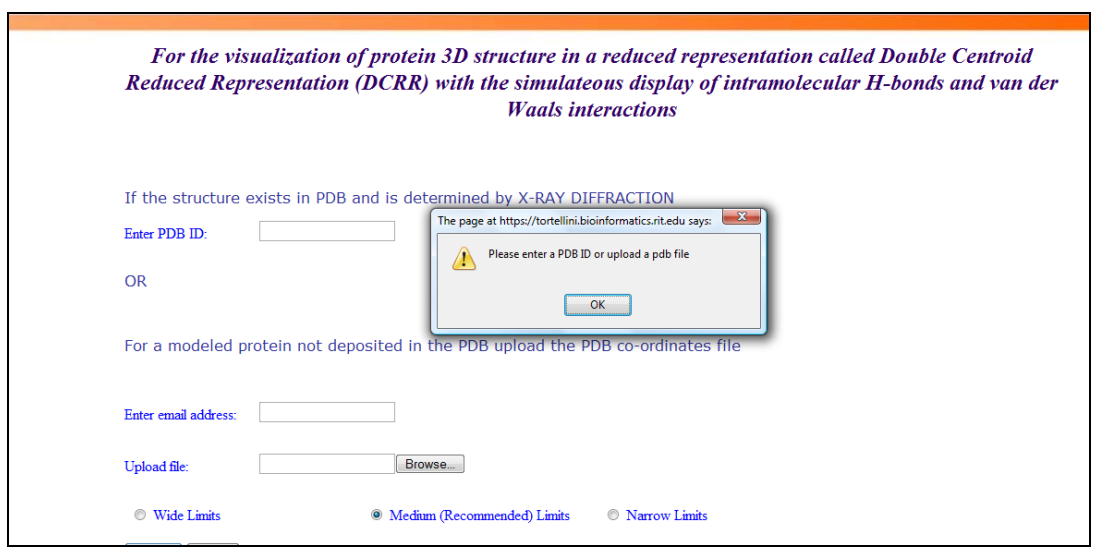

*Figure 10: Alert warnings in the web page*

Server-side validation involves a more complicated series of steps to validate the format of the uploaded file as well as the email address given. It prompts the user in case an invalid email address is typed. It rejects the file if it is not the right file type (i.e. .txt) and if the file is greater than 1MB. Only after making these changes it saves a copy of the file to the server and calls the shell script. It also alerts the user that processing will take time. After processing the browser displays the link containing the result to the browser as well as emails the results to the user. The uploaded files will be saved in a directory separate from the PDB structures to ensure the sanctity of the PDB structures. The modeled structure directory will be cleared regularly so it is advised that the users download their files as soon as they get the email.

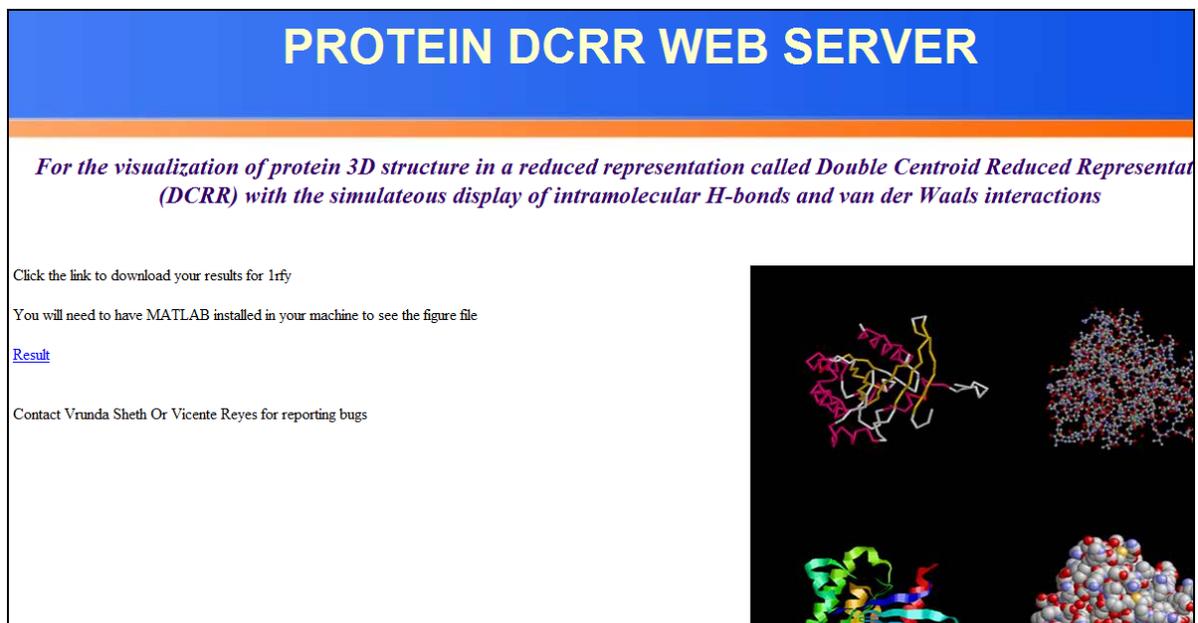

*Figure11a: Result page of protein DCRR Web Server when a PDB id is entered*

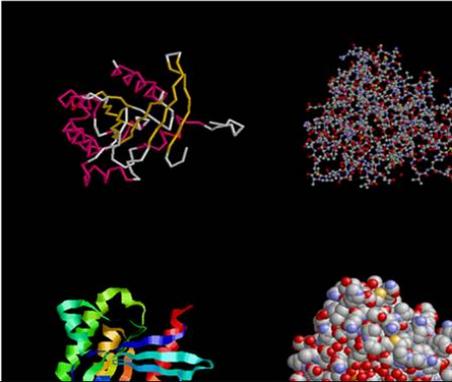

*Figure 11b: Result page of a protein DCRR Web Server when a file is uploaded*

## 4. Database creation

MySQL was used to create a database of reduced structures. The database has 3 tables, protein_wide, protein_recommended and protein_narrow for the 3 range of H-bonds and van der Waals interactions. These tables have the same structure as shown below

```
mysql> desc proteins;
+-------+--------------+------+-----+---------+-------+
| Field | Type         | Null | Key | Default | Extra |
+-------+--------------+------+-----+---------+-------+
| id    | int(11)      | NO   | PRI | NULL    |       |
| name  | varchar(4)   | NO   |     | NULL    |       |
| range | varchar(15)  | NO   |     | NULL    |       |
| link  | varchar(100) | YES  |     | NULL    |       |
+-------+--------------+------+-----+---------+-------+
4 rows in set (0.00 sec)
```

*Table 2: Description of the tables in the database*

The first column is an integer id for the protein. The second column is the name of the protein, only 4 letter strings allowed. The third column is the range. The fourth column is the link. The link points to the folder which contains the zipped file output of the web server. The zipped file has the DCRR co-ordinates of the protein, the coordinates for the H-bonds, van der Waals interactions, and ligands and water molecules as well as the MATLAB image of the protein in a reduced representation.

A separate file contains configuration instructions needed to connect to the database. The database contains only x-ray crystallographic structures in the PDB (i.e., NMR structures excluded at this time due to the high redundancy of the ensemble). The user-uploaded structures will be stored and processed in a separate directory from the database and will be cleaned up periodically.

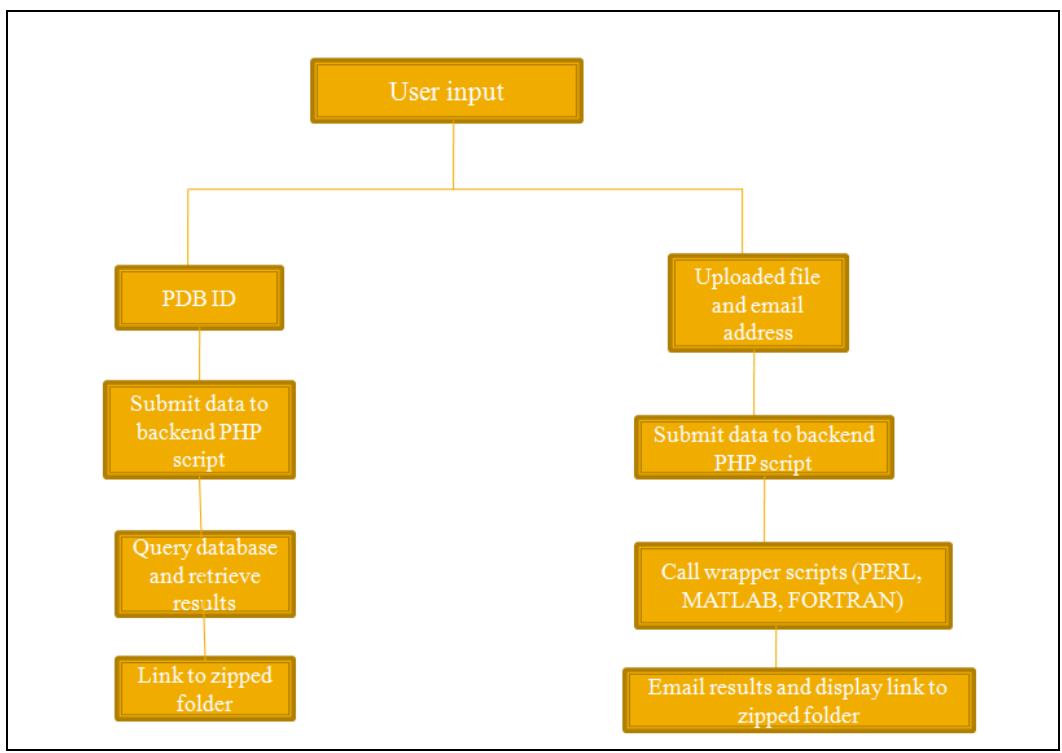

*Figure 12: Flow chart of the Web Server steps*

Figure 11 illustrates different ways in which the user information gets processed in the DCRR Web Server. When a PDB id entered, it is used to query the database and the link is sent back to the user. If a file is uploaded, it is checked for format consistency and then saved on the server. The DCRR processing scripts are called to convert the file to a reduced representation and then MATLAB generates an image of the reduced protein.

# Chapter 4

Results

We created a test dataset which consisted of 4 proteins, one each from the major SCOP family classes: all-α protein, all-β protein, α+β protein, and α/β protein. The reason for choosing a protein from each of the family class was to make sure all different sizes, and properties of the proteins are considered in the optimization steps such as H-bond optimization, and van der Waals distance optimization etc. The four proteins considered are as follows:

## 1. all-α protein -- 1rfy

This is a transcriptional repressor protein (traM). It has 2 α-helical chains. The source organism is *Agrobacterium tumefaciens*.

<http://www.rcsb.org/pdb/explore/explore.do?structureId=1RFY>.

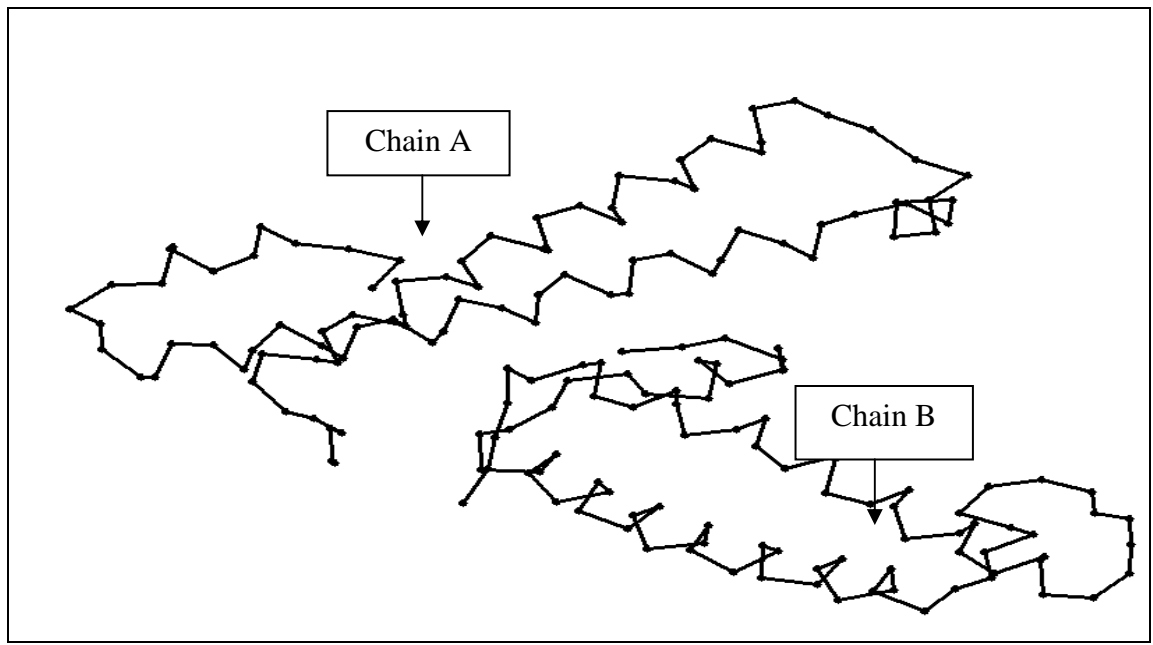

*Figure 13: Image of backbone centroids of 1rfy*

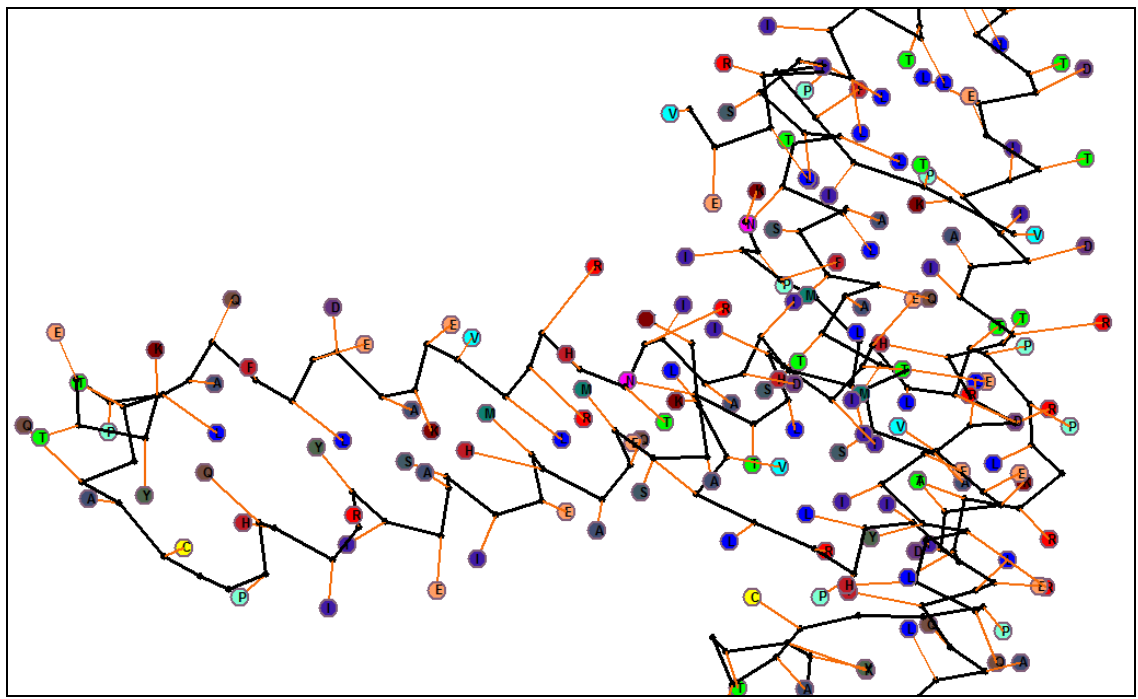

*Figure 14: This image shows backbone centroids and the side-chain centroids of 1rfy.*

This image shows the backbone centroids of 1rfy in black as seen in the previous image. The side-chain centroids of 1rfy are shown as spheres extending from the backbone centroids. Each side-chain centroid is colored according to polarity and hydrophobicity. Each side-chain centroid is also labeled with its single letter code. The secondary structure is clearly visible in this representation. As each side-chain centroid is labeled and color-coded, it becomes easier for the user to determine its identity unlike other user interfaces wherein users need to click each point to find out the amino acid at that point.

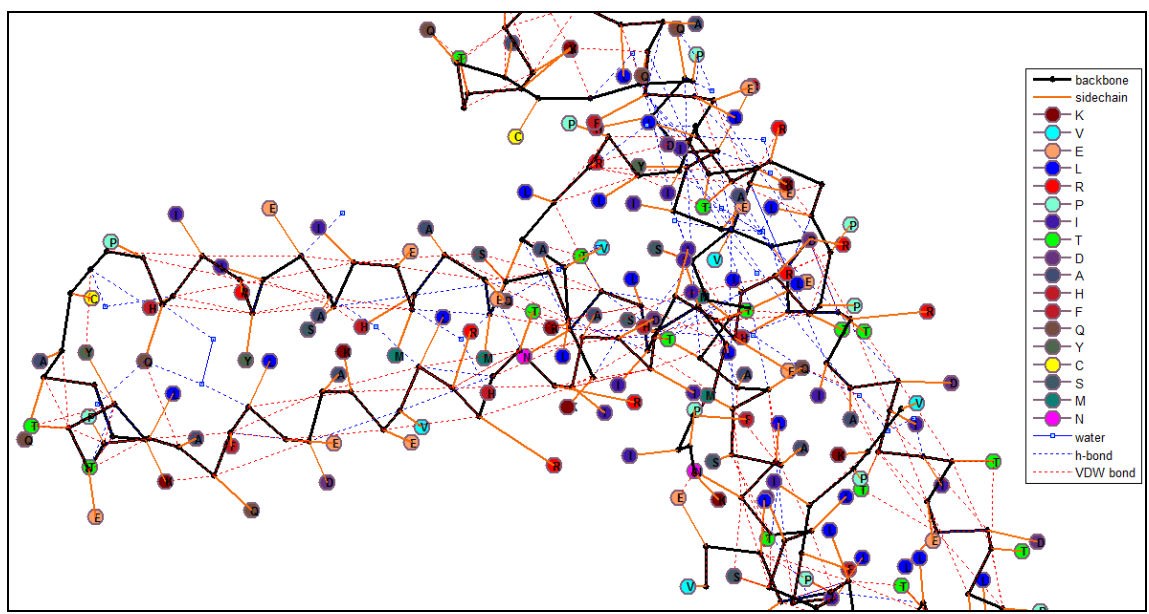

*Figure 15: DCRR of 1rfy along with H-bonds, Van der Waals interactions, ligands and bound ordered water molecules*

This is the entire DCRR image of 1rfy. It shows the backbone and the side-chain centroids as well as H-bonds and van der Waals interaction. The H-bonds are shown in blue, and van der Waals interaction is shown in red. The smaller blue squares show the bound ordered water molecules. The visualization script was programmed to output a legend for easy identification of colors for each amino acid. The MATLAB output file produced is called output.fig and can be viewed using MATLAB. The image can be rotated and translated as well as zoomed in and out.

## 2. all –β protein -1rie

1rie is a bovine heart mitochondrial complex. It has iron binding activity and an iron binding domain. It helps in electron transfer. It has 1 parallel β chain. The ligand is FeS.
<http://www.rcsb.org/pdb/explore/explore.do?structureId=1RIE>.

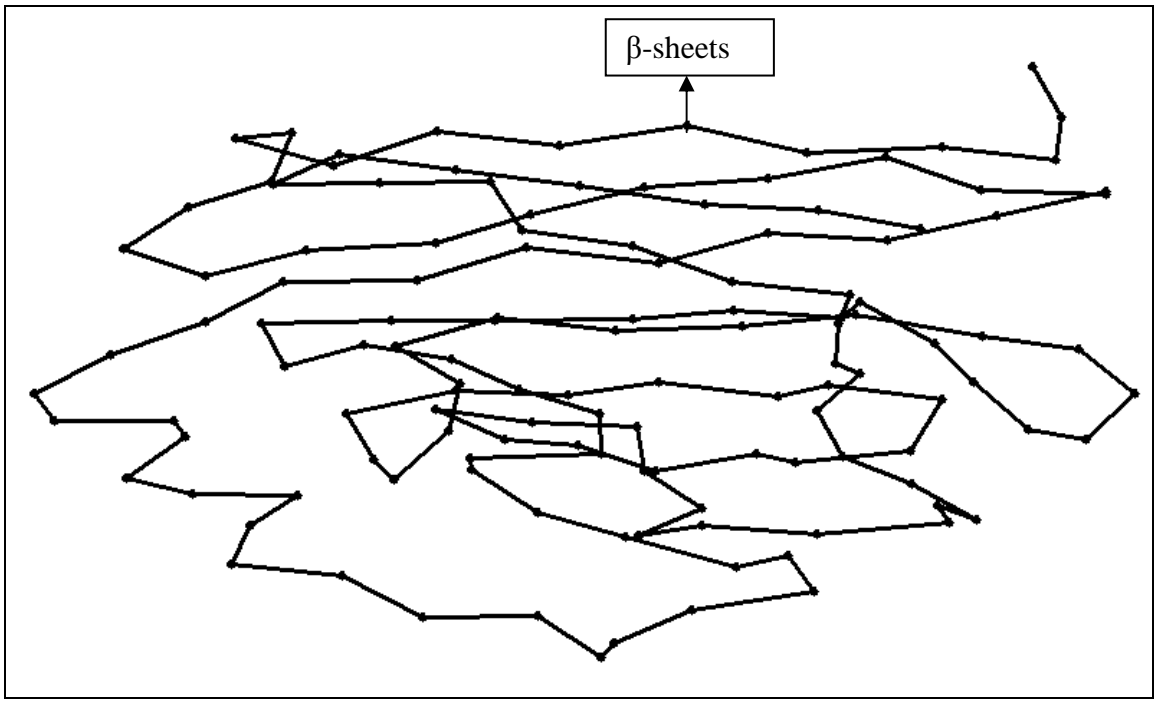

*Figure16: Image of the backbone centroids of 1rie*

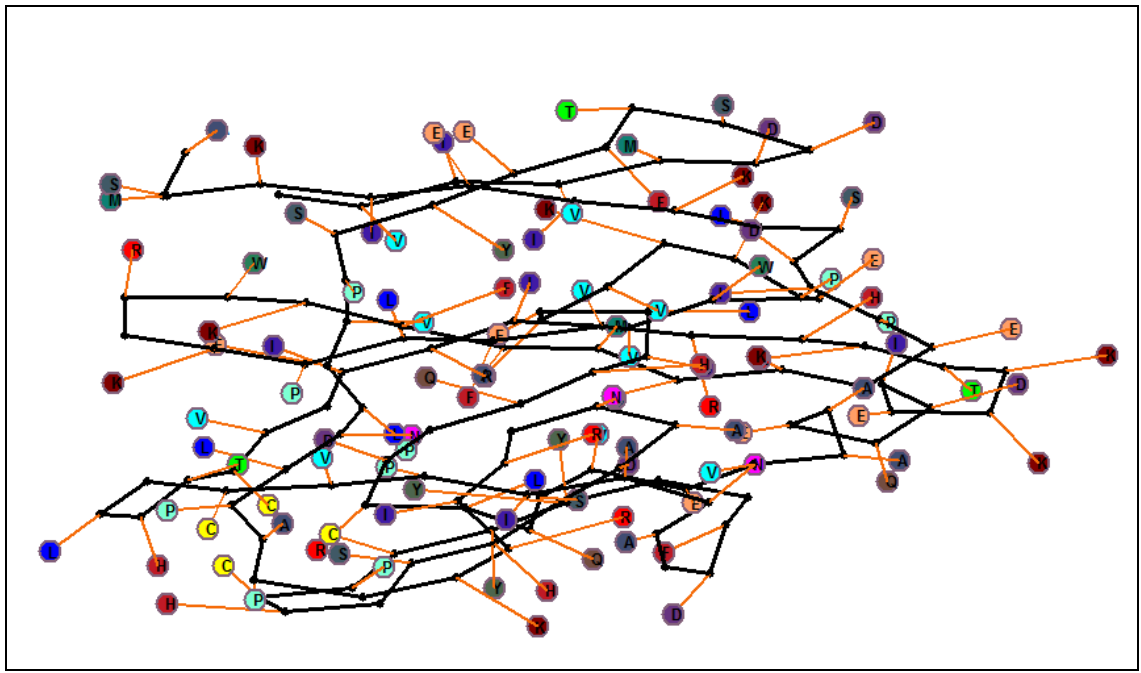

*Figure 17: Backbone and the side-chain centroids of 1rie*

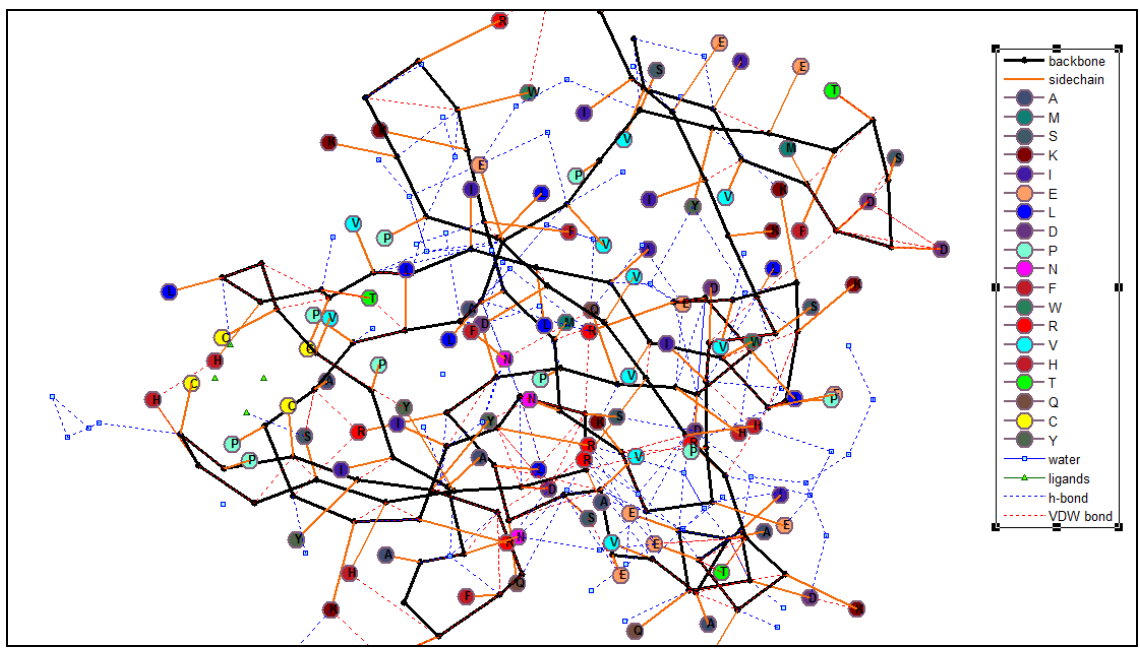

*Figure 18: DCRR image of 1rie with the H-bonds, Van der Waals interaction and the ligands and bound ordered water molecules.*

## 3. α+β protein --1hmk

1 hmk is a goat α –lactalbumin. It has lactose synthetase activity and a calcium binding site. It has 1 chain with 37% is helical and 11% is β-sheets.

< http://www.rcsb.org/pdb/explore/explore.do?structureId=1HMK>.

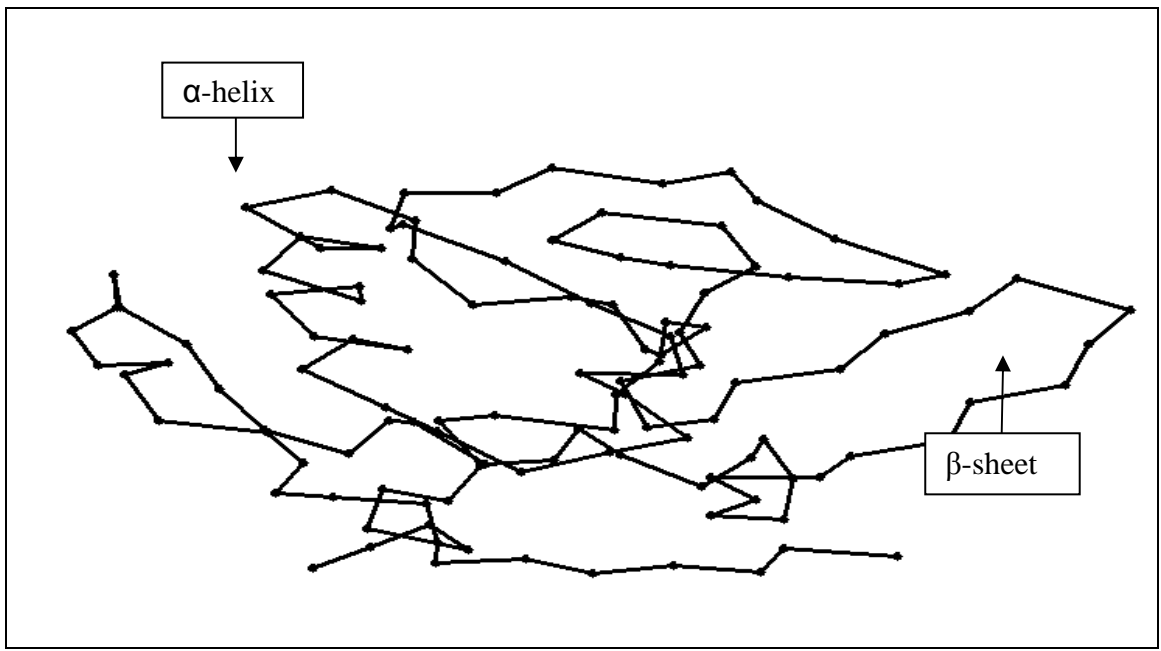

*Figure 19: Backbone centroids of 1hmk*

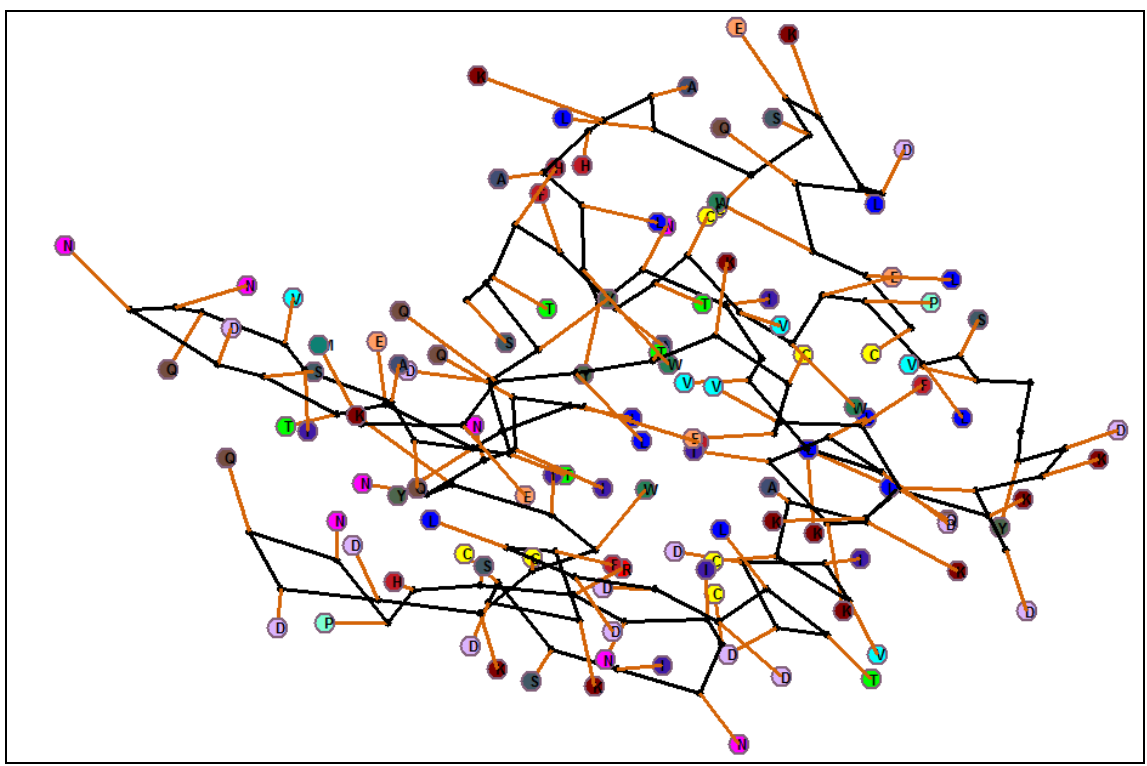

*Figure 20: Backbone and side chain centroids of 1hmk*

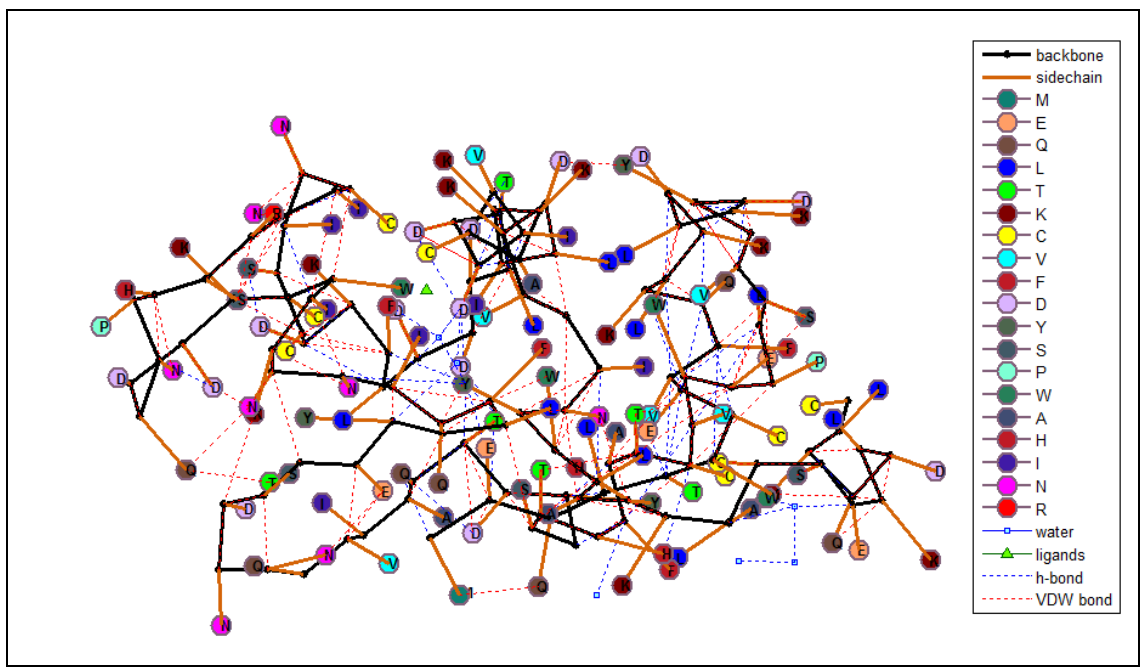

*Figure 21: DCRR image of 1hmk including H-bonds, Van der Waals interactions, and the ligands and bound ordered water molecules.*

This is the image of 1hmk in DCRR. The backbone and the side-chain centroids along with the H-bonds and the van der Waals interactions are clearly visible. The H-bonds formed between the protein and the bound ordered water molecules are also clearly visible. 1hmk has a calcium binding domain. The calcium ion and the binding site of the ion are also clearly seen in DCRR. Calcium is shown as a green triangle and labeled in the legend as a ligand.

### 4. α/β protein—4ovo.ent

This is an ovomucoid inhibitor protein obtained from Japanese quail. It is a single chain protein with 56 residues. <http://www.rcsb.org/pdb/explore/explore.do?structureId=4OVO>.

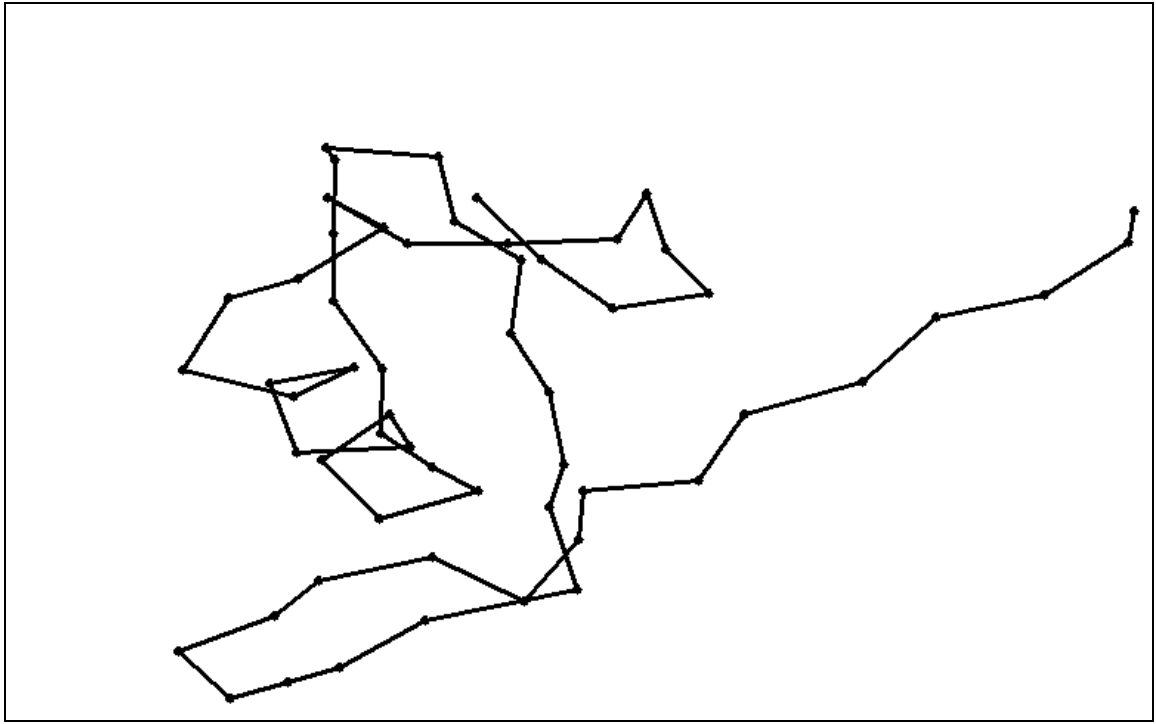

*Figure 22: Backbone centroids of 4ovo*

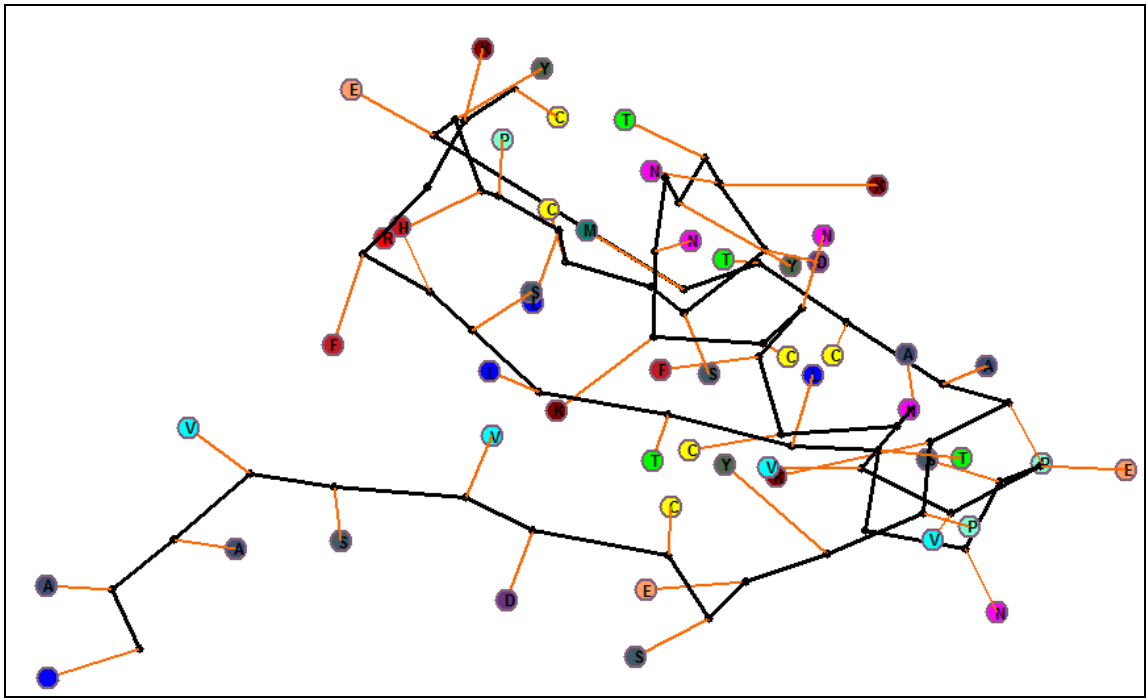

*Figure 23: Backbone and side-chain centroids of 4ovo*

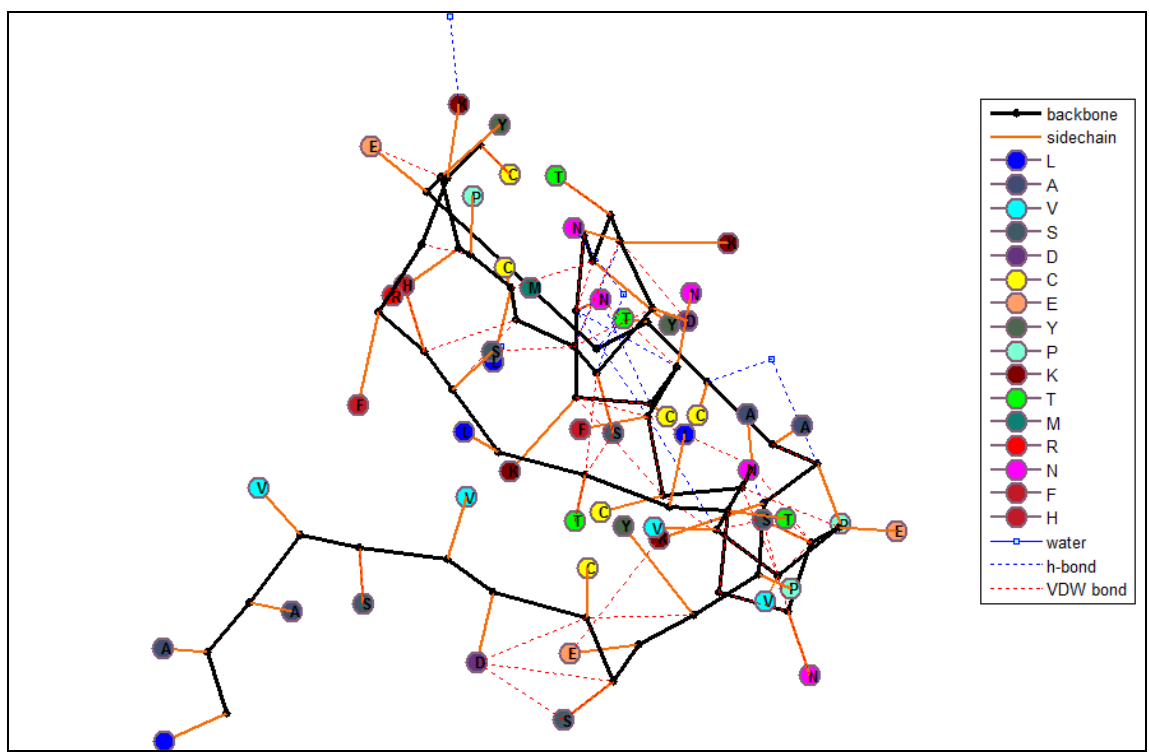

*Figure 24: DCRR of the backbone and the side-chain centroid of 4ovo along with the H-bonds, Van der Waals interactions, and the bound ordered water molecules.*

## 5. Result statistics

Currently there are about 60,000 structures in the PDB and new structures get added every Wednesday. A copy of the PDB database was downloaded onto our local server on April 24[th] 2009. Of the total structures, 41,465 were protein structures determined by x-ray crystallography. All the structures are passed through DCRR pipeline to get the reduced representation. The calculations were done for wide, narrow and recommended limits. DCRR for each of the protein structure was then deposited in the database. Of the 41,465 structures 3,700 structures had format inconsistency and these structures are not deposited in the database. It took about 3 months to do all the calculations and generate MATLAB images for the PDB structures for the 3 limits (narrow, wide and recommended).

Depending on the size of the protein the protein DCRR calculations can take anywhere between 5-50 minutes. The bottleneck step is the nearest neighbor calculation where the neighbor for every atom in a protein is calculated. Hence to prevent users from spending time doing all the calculations on the fly we developed the DCRR database wherein users could just query these structures. However for a modeled protein uploaded on the website, the calculations are done on the fly and hence can take some time.

# Chapter 6

Challenges and Conclusion

**1. Conclusion**

The idea for DCRR came from the motivation that we needed a simplified protein structure yet one that captures the biochemical properties of the molecule as well. Using the AAR representation for molecular dynamics and protein-protein interaction studies is becoming a daunting task merely due to the computational power required to analyze the sheer number of data points in a protein. However our definition of a simplified structure was a structure without loss of information. Hence, DCRR, a new method of protein structure representation was developed. DCRR allowed simplifying the protein structure information as each amino acid was represented as 2 data points, the centroid of the backbone and the centroid of the side chain reducing the overall number of data points by 70%. Since DCRR represents both the backbone and the side-chain centroids the loss of structural and biochemical information is minimal.

The MATLAB visualization tool has been developed to visualize the DCRR of the protein. To our best knowledge, this is the first visualization tool for a reduced representation of a protein. MATLAB is a good visualization interface allowing users to rotate, zoom in, zoom out and translate the protein for a better 3D view. The DCRR visualization tool allows simultaneous display of the secondary structure of the protein as well as the H-bonds and the van der Waals interactions. DCRR also shows all the ligands of the protein as well as the bound ordered water molecules. DCRR provides a good view of the ligand binding site as well as the protein water and the protein ligand interactions. The visualization script has been programmed to simultaneously display a legend for easy identification of amino acids and ligands. Unlike the other visualization tools, in DCRR we can easily identify all the amino acids, as each side chain centroid is colored and labeled according to its polarity and hydrophobicity. This is more user friendly as compared to the other visualization tools where user needs to click on each point to identify the amino acid.

The Protein DCRR Web Server has been developed as a complementary site to the PDB. The protein DCRR Web Server allows users to have easy access to reduced representation of the structures in the PDB as well as modeled proteins not deposited in the PDB. The backend of the Web Server is a protein DCRR database which has the reduced representation of all the protein in wide, narrow and reduced representation.

Apart from easy visualization there are many benefits of reduced representation. Reduced representation can be used instead of AAR for molecular dynamics studies. DCRR can also be used for pharmacophore modeling and protein-protein interaction site modeling. The ease of handling a DCRR structure makes it very useful for proteomics structure modeling applications.

## 2. Challenges

This project was a challenging project providing ample opportunity to acquire new skill sets. I was able to learn languages such as PHP and HTML and tools like MATLAB. One of the challenges associated with the project was the inconsistency in the PDB structures. Many of the structures in the PDB deviate from the standard PDB format causing many of the preprocessing scripts to break. This really slowed down the process of converting all the structures in the PDB to a reduced representation. There is a need to bring more consistency to the PDB format. One such method could be to convert the PDB format file to a tab delimited file format where each column represents a value. This could bring more standardization in the PDB allowing easy scripts to be written to process the structure files.

## 3. Future Directions

The manuscript for this work is in preparation and expected to publish 2010. This project serves as a foundation for several other projects in the future. This project can be expanded to include DCRR of DNA and RNA as well (in process), in which case it would more aptly be designated "TCRR" -- 'triple centroid reduced representation' -- where each nucleotide is represented as three centroids, namely the base, the sugar and the phosphate centroids. This project also serves as a starting point for making three different databases α-helix, β-sheet and the loop database in reduced representation for protein secondary structure association studies. Going forward, the DCRR visualization interface can be programmed into other common visualization tools like pymol and rasmol as plug-ins. DCRR can be extensively used for protein visualization as well as for modeling purposes. Another important area where DCRR can be highly beneficial is in protein molecular dynamics simulations. Finally, the proteins in DCRR can be of instructional value in the classroom and other pedagogical settings.

# Appendix

MATLAB visualization code

```
% Author name: Vrunda Sheth
% 01/27/09
% To plot the reduced representation of protein with intramolecular
% interactions.
clc;
clear all;
hold off;
hold off;

set(gcf,'color','white');
fid=fopen('protein_bbn.cm.dat');
c=textscan(fid,'%*s %d %s %s %s %s %f %f %f %*f %f %*s');

bbname=c{3};
xbb=c{6};
ybb=c{7};
zbb=c{8};

%plot3(c{6},c{7},c{8},'Linestyle','-','Color',[0 0
0],'LineWidth',2.5,'Marker','*','MarkerSize',5,'MarkerEdgeColor',[1 0.6
0.4]);
fid_tem=fopen('chainid.dat');
te=textscan(fid_tem,'%d %s');
breakPoints=te{1};
disp(breakPoints);
iterator=numel(breakPoints);
breakPoints(iterator)=breakPoints(iterator)+1;
disp(iterator);
hold on
plot3(c{6}(1:(breakPoints(2)-1)),c{7}(1:(breakPoints(2)-
1)),c{8}(1:(breakPoints(2)-1)),'Linestyle','-','Color',[0 0
0],'LineWidth',2.5,'Marker','*','MarkerSize',5,'MarkerEdgeColor',[0 0 0]);
for it= 2:(iterator-1)
    hPlot=plot3(c{6}(breakPoints(it):(breakPoints(it+1)-
1)),c{7}(breakPoints(it):(breakPoints(it+1)-
1)),c{8}(breakPoints(it):(breakPoints(it+1)-1)),'Linestyle','-','Color',[0 0
0],'LineWidth',2.5,'Marker','*','MarkerSize',5,'MarkerEdgeColor',[0 0 0]);

set(get(get(hPlot,'Annotation'),'LegendInformation'),'IconDisplayStyle','off'
);
end
hold on

hash=java.util.Hashtable;
hash.put('G',([1 0 0]));
hash.put('A',([0.25 0.3 0.45]));
hash.put('P',([0.49 1 0.83]));
hash.put('F',([0.75 0.1 0.15]));
hash.put('V',([0 1 1]));
```

```matlab
        hash.put('L',([0 0 1]));
        hash.put('I',([0.25 0.1 0.65]));
        hash.put('M',([0.05 0.5 0.45]));
        hash.put('C',([1 1 0]));
        hash.put('S',([0.25 0.35 0.4]));
        hash.put('T',([0 1 0]));
        hash.put('D',([0.4 0.2 0.5]));
        hash.put('Q',([0.45 0.3 0.25]));
        hash.put('H',([0.75 0.1 0.15]));
        hash.put('W',([0.15 0.5 0.35]));
        hash.put('Y',([0.3 0.4 0.3]));
        hash.put('N',([1 0 1]));
        hash.put('E',([1 0.62 0.4]));
        hash.put('R',([1 0 0]));
        hash.put('K',([0.5 0 0]));
        
        fid1=fopen('protein_sdc.cm.dat');
        d=textscan(fid1,'%*s %d %s %s %s %s %f %f %f %*f %*f %*s');
        scname=d{3};
        xsc=d{6};
        ysc=d{7};
        zsc=d{8};
        t=strncmp (bbname,'G',1);
        
        % to get all non-zero idndexes from t
        index=find(t);
        xbb=xbb(t~=1);
        ybb=ybb(t~=1);
        zbb=zbb(t~=1);
        
        
        
        len=numel(xbb);
        X=[xbb(1),xsc(1)];
        Y=[ybb(1),ysc(1)];
        Z=[zbb(1),zsc(1)];
        %line(X,Y,Z,'Linestyle','-','Color',[0.4 1 0],'LineWidth',2.5);
        line(X,Y,Z,'Linestyle','-','Color',[0.9608 0.4235 0.03922],'LineWidth',2.2);
        text((xsc(1)+0.1),(ysc(1)+0.1),(zsc(1)+0.1),scname(1),'Fontsize',9,'FontWeigh
        t','bold');
        
        myvector='X';
        
        for k=1:len
            % to find if a string exists in a vector
            temp=char(scname(k));
            incidence=findstr(myvector,temp);
              if (numel(incidence)==0)
                  myvector=strcat(myvector,scname(k));
                  myvector=char(myvector);
                  plot3(xsc(k),ysc(k),zsc(k),'LineWidth',2,'Color',[0.5137 0.3804
        0.4824],'Marker','o','MarkerSize',12,'MarkerFaceColor',hash.get(char(scname(k
        ))));
              else
                    sLine=plot3(xsc(k),ysc(k),zsc(k),'LineWidth',2,'Color',[0.5137
        0.3804
```

```matlab
0.4824],'Marker','o','MarkerSize',12,'MarkerFaceColor',hash.get(char(scname(k))));
set(get(get(sLine,'Annotation'),'LegendInformation'),'IconDisplayStyle','off');
      end;
end

for k=2:len
X=[xbb(k),xsc(k)];
Y=[ybb(k),ysc(k)];
Z=[zbb(k),zsc(k)];
%hLine=line(X,Y,Z,'Linestyle','-','Color',[0.4 1 0],'LineWidth',2.5);
hLine=line(X,Y,Z,'Linestyle','-','Color',[0.9608 0.4235 0.03922],'LineWidth',2.2);
set(get(get(hLine,'Annotation'),'LegendInformation'),'IconDisplayStyle','off');
text((xsc(k)-0.05),(ysc(k)-0.01),(zsc(k)-0.01),scname(k),'Fontsize',9,'FontWeight','bold');
uni=unique(scname);
%plot3(xsc(k),ysc(k),zsc(k),'LineWidth',1,'Color',[0.5137 0.3804 0.4824],'Marker','o','MarkerSize',13,'MarkerFaceColor',hash.get(char(scname(k))));
end

myvector=myvector(2:end);
uni=unique(scname);
tot=numel(myvector);
 final=zeros(1,tot);
 for tem =1:tot
     ano(tem)=cellstr(myvector(tem));
 end

ano=(ano)';

% legend(legend_names);

fid_het=fopen('wat.aar.dat');
het=textscan(fid_het,'%*s %*d %*s %*s %*s %*d %f %f %f %*f %*f %*s');
het_len=numel(het{1});
if(het_len >0)

plot3(het{1}(1),het{2}(1),het{3}(1),'LineWidth',1,'Marker','s','MarkerSize',3,'MarkerFaceColor',[0.6 1 1]);
    for jj= 2:het_len

hPlot=plot3(het{1}(jj),het{2}(jj),het{3}(jj),'LineWidth',1,'Marker','s','MarkerSize',3,'MarkerFaceColor',[0.6 1 1]);

set(get(get(hPlot,'Annotation'),'LegendInformation'),'IconDisplayStyle','off');
    end
end
```

```
    fid_lig=fopen('lig.aar.dat');
    het_lig=textscan(fid_lig,'%*s %*d %*s %*s %*s %*d %f %f %f %*f %*f %*s');
    het_lig_len=numel(het_lig{1});
    if (het_lig_len > 0 )
        plot3(het_lig{1}(1),het_lig{2}(1),het_lig{3}(1),'LineWidth',1,'Color',[0
0.4 0],'Marker','^','MarkerSize',5.5,'MarkerFaceColor',[0.4 1 0]);
        for jj= 2:het_lig_len

hPlot=plot3(het_lig{1}(jj),het_lig{2}(jj),het_lig{3}(jj),'LineWidth',1,'Color
',[0 0.4 0],'Marker','^','MarkerSize',5.5,'MarkerFaceColor',[0.4 1 0]);

set(get(get(hPlot,'Annotation'),'LegendInformation'),'IconDisplayStyle','off'
);
        end
    end

    fid2=fopen('final_hbo.dat');
    e=textscan(fid2,'%*d %*s %*s %f %f %f %*d %*s %*s %f %f %f' );
    X=[e{1}(1),e{4}(1)];
    Y=[e{2}(1),e{5}(1)];
    Z=[e{3}(1),e{6}(1)];
    line(X,Y,Z,'Linestyle',':','Color','b','LineWidth',1.25);
    elen=length (e{1});
    for s=2:elen
        X=[e{1}(s),e{4}(s)];
        Y=[e{2}(s),e{5}(s)];
        Z=[e{3}(s),e{6}(s)];
        hLine=line(X,Y,Z,'Linestyle',':','Color','b','LineWidth',1.25);

set(get(get(hLine,'Annotation'),'LegendInformation'),'IconDisplayStyle','off'
);
    end

    fid3=fopen('final_vdw.dat');
    f=textscan(fid3,'%*d %*s %*s %f %f %f %*d %*s %*s %f %f %f' );
    X=[f{1}(1),f{4}(1)];
    Y=[f{2}(1),f{5}(1)];
    Z=[f{3}(1),f{6}(1)];
    line(X,Y,Z,'Linestyle',':','Color','r','LineWidth',1.25);
    flen=length (f{1});
    for s=2:flen
        X=[f{1}(s),f{4}(s)];
        Y=[f{2}(s),f{5}(s)];
        Z=[f{3}(s),f{6}(s)];
        hLine=line(X,Y,Z,'Linestyle',':','Color','r','LineWidth',1.25);

set(get(get(hLine,'Annotation'),'LegendInformation'),'IconDisplayStyle','off'
);
    end
    if (het_len > 0 && het_lig_len >0)
         legend_names=['backbone'; 'sidechain'; ano; 'water' ;'ligands';'h-
bond';'VDW bond'];
    end
    if (het_len > 0 && het_lig_len <=0 )
```

```
    legend_names=['backbone'; 'sidechain'; ano; 'water' ;'h-bond';'VDW
bond'];
end
if (het_len <= 0 && het_lig_len <=0)
     legend_names=['backbone'; 'sidechain'; ano;'h-bond';'VDW bond'];
end
if (het_len <= 0 && het_lig_len > 0)
     legend_names=['backbone'; 'sidechain'; ano;'ligands';'h-bond';'VDW
bond'];
end

legend(legend_names);
%legend;
axis equal;
axis off;
hgsave('output');
exit
```

Wrapper script

```
#!/bin/bash

grep "^ATOM" pdb > protein.aar

grep "^HETATM" pdb > hetatm.aar

grep "HOH" hetatm.aar > water.aar

cp hetatm.aar filei

/home/vns4483/totalScripts/ligand_reduction.x

cp fileo hetatm.aar

/home/vns4483/totalScripts/remove_wat.pl water.aar > water1.aar

grep -v "HOH" hetatm.aar > ligand.aar

sed -e '1,$s/^ATOM/PRT /g' protein.aar > prt.aar

sed -e '1,$s/^HETATM/WAT   /g' water1.aar > wat.aar.dat

sed -e '1,$s/^HETATM/LIG   /g' ligand.aar > lig.aar.dat

cat prt.aar lig.aar.dat wat.aar.dat > total.aar

sort +1 -2 -n total.aar > plw.aar

mv plw.aar filei

/home/vns4483/totalScripts/bfactor_reduction.x
```

```
mv fileo plw.aar

cp prt.aar filei

/home/vns4483/totalScripts/ResidueNumberReduction.x

mv fileo prt.aar

cp plw.aar filei

/home/vns4483/totalScripts/ResidueNumberReduction.x

mv fileo plw.aar.reduced

cp plw.aar.reduced plw_copy.aar.reduced

/home/vns4483/totalScripts/get_backbone.x

/home/vns4483/totalScripts/get_sidechain.x

echo "PRT       0   X    XXX X   1      -15.809  -1.266  48.400  1.0  27.36     X" >> protein.bbn

echo "PRT       0   X    XXX X   1      -15.809  -1.266  48.400  1.0  27.36     X" >> protein.sdc

/home/vns4483/totalScripts/res2cm_bbn.x

/home/vns4483/totalScripts/res2cm_sdc.x

/home/vns4483/totalScripts/make_single_letter.pl protein_bbn.cm > protein_bbn.cm.dat

/home/vns4483/totalScripts/make_single_letter.pl protein_sdc.cm > protein_sdc.cm.dat

cp protein_bbn.cm filea

cp filea fileb

/home/vns4483/totalScripts/find_chainIDs.x

wc -l protein_bbn.cm >line1

cat fileo.dat line1 > chainid.dat

cat protein_bbn.cm protein_sdc.cm > preDC

cp preDC filei

/home/vns4483/totalScripts/bfactor_reduction.x

mv fileo preDC
```

```
sort +1 -2 -n preDC > prot.dcrr
/home/vns4483/totalScripts/nrst_ngbr.x
/home/vns4483/totalScripts/find_Hbonds.x
cp protein.hbo filei
/home/vns4483/totalScripts/remove_PRT_PRT.x
rm filei
mv fileo inter_prt_lig_wat.hbo
cp inter_prt_lig_wat.hbo filei
/home/vns4483/totalScripts/interM12DCRR.x
chmod ugo+x filex
./filex > inter_1_hbo
/home/vns4483/totalScripts/make_2_files.pl inter_1_hbo
/home/vns4483/totalScripts/cleaning_up.pl one_side oth_side
/home/vns4483/totalScripts/unique.pl outfile > inter_2_hbo
/home/vns4483/totalScripts/buffer.pl inter_2_hbo > temp
mv temp inter_2_hbo
/home/vns4483/totalScripts/remove_inversions.pl inter_2_hbo > inter_final_hbo
grep -v 'LIG' protein.hbo | grep -v 'WAT' > intra_hbo
cp intra_hbo filei
/home/vns4483/totalScripts/intraM12DCRR.x
chmod ugo+x filex
./filex > intra_1_hbo
/home/vns4483/totalScripts/make_2_files.pl intra_1_hbo
/home/vns4483/totalScripts/cleaning_up.pl one_side oth_side
/home/vns4483/totalScripts/unique.pl outfile > intra_2_hbo
/home/vns4483/totalScripts/buffer.pl intra_2_hbo > temp
mv temp intra_2_hbo
```

```
/home/vns4483/totalScripts/remove_inversions.pl intra_2_hbo > intra_final_hbo

cat inter_final_hbo intra_final_hbo >final_hbo.dat

/home/vns4483/totalScripts/find_VDW.x

rm filew filex filey filez

cp protein.vdw filei

/home/vns4483/totalScripts/remove_PRT_PRT.x

rm filei

mv fileo inter_prt_lig_wat.vdw

cp inter_prt_lig_wat.vdw filei

/home/vns4483/totalScripts/interM12DCRR.x

chmod ugo+x filex

./filex > inter_1_vdw

/home/vns4483/totalScripts/make_2_files.pl inter_1_vdw

/home/vns4483/totalScripts/cleaning_up.pl one_side oth_side

/home/vns4483/totalScripts/unique.pl outfile > inter_2_vdw

/home/vns4483/totalScripts/buffer.pl inter_2_vdw > temp

mv temp inter_2_vdw

/home/vns4483/totalScripts/remove_inversions.pl inter_2_vdw > inter_final_vdw

grep -v 'LIG' protein.hbo | grep -v 'WAT' > intra_vdw

cp intra_vdw filei

/home/vns4483/totalScripts/intraM12DCRR.x

chmod ugo+x filex

./filex > intra_1_vdw

/home/vns4483/totalScripts/make_2_files.pl intra_1_vdw

/home/vns4483/totalScripts/cleaning_up.pl one_side oth_side

/home/vns4483/totalScripts/unique.pl outfile > intra_2_vdw
```

```
/home/vns4483/totalScripts/buffer.pl intra_2_vdw > temp

mv temp intra_2_vdw

/home/vns4483/totalScripts/remove_inversions.pl intra_2_vdw >
intra_final_vdw

cat inter_final_vdw intra_final_vdw >final_vdw.dat

matlab -nodesktop -nosplash -r colde

mkdir dcrr

mv prot.dcrr dcrr/

mv final_vdw.dat dcrr/

mv final_hbo.dat dcrr/

mv output.fig dcrr/

mv lig.aar.dat dcrr/

mv wat.aar.dat dcrr/

tar -zcvf dcrr.tar.gz dcrr

rm chainid.dat filea fileb filei fileo fileo.dat fileu filex hetatm.aar
inter_1_hbo inter_1_vdw inter_2_hbo inter_2_vdw inter_final_hbo
inter_final_vdw intra_hbo intra_vdw ligand.aar line1 one_side oth_side
outfile pdb plw.aar plw.aar.reduced plw_copy.aar.reduced preDC protein.aar
protein.bbn protein_bbn.cm protein.hbo protein.nnb protein.sdc protein.vdw
prt.aar total.aar water1.aar water.aar inter_prt_lig_wat.hbo
inter_prt_lig_wat.vdw intra_1_hbo intra_1_vdw intra_2_hbo intra_2_vdw
intra_final_hbo intra_final_vdw protein_bbn.cm.dat protein_sdc.cm
protein_sdc.cm.dat

exit
```

Php Code

```
<!DOCTYPE html PUBLIC "-//W3C//DTD XHTML 1.0 Transitional//EN"
        "http://www.w3.org/TR/xhtml1/DTD/xhtml1-transitional.dtd">
<html xmlns="http://www.w3.org/1999/xhtml">

<head>
        <title>PROTEIN DCRR WEB SERVER</title>
<style type="text/css">

body
{background-image:url(page-background.jpg);background-repeat:no-
repeat;background-position:100% 7%;
```

```
    margin-left: 10%; margin-right: 10%; color: blue;}
#loading {
        width: 200px;
        height: 100px;
        background-color: #c0c0c0;
        position: absolute;
        left: 50%;
        top: 50%;
        margin-top: -50px;
        margin-left: -100px;
        text-align: center;
}
##############################################################################
# The following section has been developed using the link below as reference.
#("Ask the CCS Guy",2009).
#<http://www.askthecssguy.com/2007/03/form_field_hints_with_css_and.html>.
##############################################################################
"<http://www.javascriptkit.com/script/script2/formfieldhints.shtml>.
dl {
        font:normal 12px/15px Times Bold;
    position: relative;
    width: 800px;
}
dt {
    clear: both;
    float:left;
    width: 140px;
    padding: 4px 0 2px 0;
    text-align: left;
}
dd {
    float: left;
    width: 250px;
    margin: 0 0 8px 0;
    padding-left: 6px;
}

/* The hint to Hide and Show */
.hint {
        font:normal 12px/14px Courier;
    display: none;
    position: absolute;
    right: 10px;
    width: 200px;
    margin-top: -4px;
    border: 1px solid #c93;
    padding: 10px 12px;
    /* to fix IE6, I can't just declare a background-color,
    I must do a bg image, too!  So I'm duplicating the pointer.gif
    image, and positioning it so that it doesn't show up
    within the box */
    background: #ffc url(pointer.gif) no-repeat -10px 5px;
}

/* The pointer image is hadded by using another span */
.hint .hint-pointer {
```

```
    position: absolute;
    left: -10px;
    top: 5px;
    width: 9px;
    height: 19px;
    background: url(pointer.gif) left top no-repeat;
}
</style>

<script type="text/javascript">
function addLoadEvent(func) {
  var oldonload = window.onload;
  if (typeof window.onload != 'function') {
    window.onload = func;
  } else {
    window.onload = function() {
      oldonload();
      func();
    }
  }
}

function prepareInputsForHints() {
        var inputs = document.getElementsByTagName("input");
        for (var i=0; i<inputs.length; i++){
                // test to see if the hint span exists first
                if (inputs[i].parentNode.getElementsByTagName("span")[0]) {
                        // the span exists!  on focus, show the hint
                        inputs[i].onfocus = function () {

        this.parentNode.getElementsByTagName("span")[0].style.display =
"inline";
                        }
                        // when the cursor moves away from the field, hide the
hint
                        inputs[i].onblur = function () {

        this.parentNode.getElementsByTagName("span")[0].style.display =
"none";
                        }
                }
        }
        // repeat the same tests as above for selects
        var selects = document.getElementsByTagName("select");
        for (var k=0; k<selects.length; k++){
                if (selects[k].parentNode.getElementsByTagName("span")[0]) {
                        selects[k].onfocus = function () {

        this.parentNode.getElementsByTagName("span")[0].style.display =
"inline";
                        }
                        selects[k].onblur = function () {

        this.parentNode.getElementsByTagName("span")[0].style.display =
"none";
                        }
                }
```

```
                }
        }
        addLoadEvent(prepareInputsForHints);
        </script> "
        ########################################################################
        #Reference ends here
        ########################################################################
        <script type="text/javascript" language="javascript">
        function validateMyForm() {
        var pid, file;
        with(window.document.form1)
        {
                pid=pdbid;
                file=uploadfile;
        }
        if((trim(pid.value)=='') && (trim(file.value)==''))
        {
                alert('Please enter a PDB ID or upload a pdb file');
                pid.focus();
                return false;
        }
        else  if((!trim(pid.value)=='') && (!trim(file.value)==''))
        {
                alert('Enter either a PDB ID of a protein or upload a file.');
                pid.focus();
                return false;
        }
        else if ((!trim(pid.value)=='') && (trim(file.value)==''))
        {
                return true;
        }
        else if ((trim(pid.value)=='') && (!trim(file.value)==''))
        {
                alert('Processing will take some time.Please wait patiently');
                file.focus();
                return true;
        }
        }
        function trim(str)
        {
           return str.replace(/^\s+|\s+$/g,'');
        }
        </script>
        </head>
        <body>
        <br>
        </br>
        <A HREF="http://www.bioinformatics.rit.edu"><img style="position: absolute;
        top: 0.5; right: 0;"  src="bioinfologo.gif" ALT="RIT Bioinformatics"/></A>
        <h1 align="center"> <font face="Arial" size="9" color="#FFFFCC"><b>PROTEIN
        DCRR WEB SERVER</b> </font> </h1>
        <br>
        </br>
        <br>
        </br>
        <p align="center"> <font face="Flat brush" size ="5" color="#330066" ><b> <i>
        For the visualization of protein 3D structure in a reduced representation
```

```html
called Double Centroid Reduced Representation (DCRR) with the simulateous
display of intramolecular H-bonds and van der Waals interactions
</i></b></font></p>
<br>
</br>
<br>
</br>
<form enctype="multipart/form-data" method="post" name="form1"
action="dbconnect.php">

        <p align="left"> <font face="Verdana" size ="3" color="#333399" >If
the structure exists in PDB and is determined by X-RAY DIFFRACTION
</font></p>
        <dl>
        <dt>
                <label for="pdbid">Enter PDB ID:</label>
        </dt>
        <dd>
                <input
                        name="pdbid"
                        id="pdbid"
                        maxLength=4
                        type="text" />
                <span class="hint">Enter the 4 letter PDB ID.<span
class="hint-pointer"> </span></span>
        </dd>
        </dl>
<br></br><br></br>
        <p align="left"> <font face="Verdana" size ="3" color="#333399"> OR
</font> </p>
        <br>
        </br>
        <p align="left"> <font face="Verdana" size ="3" color="#333399" >For a
modeled protein not deposited in the PDB upload the PDB co-ordinates file
</font></p>
        <br>
        </br>
        <dl>
         <dt>
                <label for="email">Enter email address:</label>
        </dt>
        <dd>
                <input
                        name="email"
                        id="email"
                        type="text" />
                <span class="hint">Enter the email address where you want the
result to be mailed too.<span class="hint-pointer"> </span></span>
        </dd>
        </dl>
<br></br><br></br>
        <dl>
        <dt>
                <label for="Uploadfile">Upload file:</label>
        </dt>
        <dd>
```

```html
                <input
                        name="uploadfile"
                        id="uploadfile"
                        type="file"
                        input type="hidden" name="MAX_FILE_SIZE" value="1000000" />
                <span class="hint">The uploaded file should be in PDB format.<span class="hint-pointer"> </span></span>
        </dd>
</dl>
<br></br><br></br>
<dl>
        <dd>
                <input type="radio" name="group1" value="Wide"/> Wide Limits<br></br>
                <span class="hint">The wide limits are 2.66 and 3.36 for H-bonds and 3.1 and 3.95 for Van der Waals interactions.<span class="hint-pointer"> </span></span>
        </dd>

        <dd>
                <input type="radio" name="group1" value="Recommended" Checked/> Recommended Limits<br></br>
                <span class="hint">The recommended limits are 2.72 and 3.22 for H-bonds and 3.20 and 3.80 for Van der Waals interactions.<span class="hint-pointer"> </span></span>
        </dd>
        <dd>
                <input type="radio" name="group1" value="Narrow" /> Narrow Limits<br></br>
                <span class="hint">The narrow limits are 2.75 and 3.0 for  H-bonds and 3.3 and 3.75 for Van der Waals interactions..<span class="hint-pointer"> </span></span>
        </dd>

</dl>

<br></br><br></br>

<input type="submit" value="Submit" onClick="return validateMyForm();"/>
<input type="reset" />
</form>
<p align="left"> <font face="Verdana" size ="2" color="#333399"> Contact Vrunda Sheth(vns4483@rit.edu) OR Vicente Reyes (vmrsbi@rit.edu)</font> </p>
<div id ='processingHook' name='processing'></div>
<div id='resultHook' name='resultHook'></div>
</body>
</html>
```

Back-end code

```
<!DOCTYPE html PUBLIC "-//W3C//DTD XHTML 1.0 Transitional//EN"
        "http://www.w3.org/TR/xhtml1/DTD/xhtml1-transitional.dtd">
<html xmlns="http://www.w3.org/1999/xhtml">
```

```html
<head>
        <title>RESULT PAGE</title>
<style type="text/css">
body
{background-image:url(page-background.jpg);background-repeat:no-
repeat;background-position:100% 7%;
}
</style>
</head>
<body>
<br>
</br>
<h1 align="center"> <font face="Arial" size="9" color="#FFFFCC"><b>PROTEIN
DCRR WEB SERVER</b> </font> </h1>
<br>
</br>
<br>
</br>
<p align="center"> <font face="Flat brush" size ="5" color="#330066" ><b> <i>
For the visualization of protein 3D structure in a reduced representation
called Double Centroid Reduced Representation (DCRR) with the simulateous
display of intramolecular H-bonds and van der Waals interactions
</i></b></font></p>
<br>
</br>
<br>
</br>

</br>
</body>
</html>
```
```php
<?php
#####################################################################
#( "Web Services Wiki", 2009).
<http://www.stanford.edu/dept/its/communications/webservices/wiki/index.php/H
ow_to_perform_error_handling_in_PHP>.
##########################################################################
#
#error_reporting(0);

"@ini_set('display_errors', 0);

ini_set('log_errors', 1);

// Destinations
define("LOG_FILE", "/home/vns4483/public_html/error.log");

// Destination types
define("DEST_EMAIL", "1");
define("DEST_LOGFILE", "3");

/* Examples */

// Send an e-mail to the administrator
#error_log("Fix me!", DEST_EMAIL, ADMIN_EMAIL);
```

```php
      // Write the error to our log file
      error_log("Error", DEST_LOGFILE, LOG_FILE);
      function my_error_handler($errno, $errstr, $errfile, $errline)
      {
        switch ($errno) {
          case E_USER_ERROR:
            // Send an e-mail to the administrator
            error_log("Error: $errstr \n Fatal error on line $errline in file
      $errfile \n", DEST_EMAIL, ADMIN_EMAIL);

            // Write the error to our log file
            error_log("Error: $errstr \n Fatal error on line $errline in file
      $errfile \n", DEST_LOGFILE, LOG_FILE);
            break;

          case E_USER_WARNING:
            // Write the error to our log file
            error_log("Warning: $errstr \n in $errfile on line $errline \n",
      DEST_LOGFILE, LOG_FILE);
            break;

          case E_USER_NOTICE:
            // Write the error to our log file
            error_log("Notice: $errstr \n in $errfile on line $errline \n",
      DEST_LOGFILE, LOG_FILE);
            break;

          default:
            // Write the error to our log file
            error_log("Unknown error [#$errno]: $errstr \n in $errfile on line
      $errline \n", DEST_LOGFILE, LOG_FILE);
            break;
        }

        return TRUE;
      }

      $old_error_handler = set_error_handler("my_error_handler");
      "
      ##############################################################################
      ##########################################
      include('helper.php');

      include('/home/vns4483/x/config.php');
      $pdbid    = input_val($_POST["pdbid"]);
      $file     = input_val($_POST["uploadfile"]);
      $range_type  = input_val($_POST["group1"]);
      $email    =input_val($_POST["email"]);

      $db1 = mysql_connect($dbhost, $dbuser, $dbpass) or die ("Unable to connect")
      ;
      $rv = mysql_select_db($dbname, $db1) or die ("Unable to select database");
      if(!$pdbid=="")
      {
      if ($range_type == 'Recommended')
      {
            $query="select * from protein_recommended where name like '$pdbid'";
```

```php
        $result=mysql_query($query);
        $num=mysql_numrows($result);
}
elseif ($range_type == 'Narrow')
{
        $query="select * from protein_narrow where name like '$pdbid'";
        $result=mysql_query($query);
        $num=mysql_numrows($result);
}
else
{
        $query="select * from protein_wide where name like '$pdbid'";
        $result=mysql_query($query);
        $num=mysql_numrows($result);
}

if ($num==0)
{
     echo "Sorry PDB ID not found <br>";
     exit;
}
$i=0;
echo "Click the link to download your results for $pdbid<br>";
echo "<br>";
echo "You will need to have MATLAB installed in your machine to see the
figure file<br>";
echo "<br>";
$out=mysql_result($result,0,"name");
$out1=mysql_result($result,0,"link");
echo "<a href='$out1'>Result</a>";
echo "<br><br>";
}
else
{

     #echo "$email";
     if (is_uploaded_file($_FILES['uploadfile']['tmp_name']))
     {
          if(!$email=="")
          {
               $bol=check_email($email);
               #echo"$bol";
               if(!$bol)
               {
                    echo "Enter a valid email address\n";
                    exit;
               }
          }
          else
          {
               echo "Please enter an email address<br>";
               exit;

          }
     }
     else
     {
```

```php
        echo "Please upload a file<br>";
        exit;
}
$filename=basename($_FILES["uploadfile"]["name"]);
$findme='ATOM';
$ext=substr($filename,strpos($filename,'.')+1);
$target_path="/home/vns4483/public_html/uploads/";
$newname = $target_path.$filename;
$dirname=substr($filename,0,4);
#$target_path="/home/vns4483/public_html/";
if(($ext=="txt") && ($_FILES["uploadfile"]["size"] < 1000000))
{
        #echo "$newname";
        if (is_uploaded_file($_FILES['uploadfile']['tmp_name']))
        {
                #echo "Displaying contents\n";
                $handle = fopen($_FILES['uploadfile']['tmp_name'], "r");
                $line1=fgets($handle);
                #echo "$line1";
                $pos1=strpos($line1,'HEADER');
                if($pos1!==0)
                {
                        echo "Not a PDB file<br>";
                        exit(1);
                }
                  else
                {
                        while(!feof($handle))
                        {
                                $line=fgets($handle);
                                $pos=strpos($line,$findme);
                                if($pos === 0 )
                                {
                                        $nf=preg_split("/\s+/",$line);
                                        $a= count($nf);
                                        #echo "$a<br>";
                                        if(sizeof($nf)== 13)
                                        {
                                                break  ;
                                        }
                                        else
                                        {
                                                echo "Not in PDB format";
                                                exit(1);
                                        }
                                }

                        }
                }
        }
        else
        {
                echo "Not uploaded the right type of file <br>";
                exit(1);
```

```php
            }

            if (!file_exists($newname))
            {
                    system("rm -r /home/vns4483/public_html/uploads/$dirname");
                    //Move uploaded file to new location                        if (move_uploaded_file($_FILES['uploadfile']['tmp_name'],$newname))
                    {
                            #$dirname=substr($filename,0,4);
                            #system("rm -r /home/vns4483/public_html/uploads/$dirname");
                            system("mkdir /home/vns4483/public_html/uploads/$dirname");
                            system("mv $newname /home/vns4483/public_html/uploads/$dirname/pdb");
                            #system("cp /home/vns4483/public_html/final_run.sh /home/vns4483/public_html/uploads/$dirname");
                            #chdir('try/$dirname');
                            system("cp /home/vns4483/public_html/colde.m /home/vns4483/public_html/uploads/$dirname");
                            #system('pwd');
                            #chdir('try/dirname');
                            #system("pwd");
                            #chdir('$dirname');
                            #system('pwd');
                            #chdir ("/home/vns4483/public_html/try/$dirname");
                            #system('pwd');
                            if ($range_type == 'Recommended')
                            {
                                    system("cp /home/vns4483/public_html/final_run.sh /home/vns4483/public_html/uploads/$dirname");
                                    chdir ("/home/vns4483/public_html/uploads/$dirname");
                                    #system('pwd');
                                    system("./final_run.sh");
                            }
                            elseif ($range_type == 'Narrow')
                            {
                                    system("cp /home/vns4483/public_html/final_run_narrow.sh /home/vns4483/public_html/uploads/$dirname");
                                    chdir ("/home/vns4483/public_html/uploads/$dirname");
                                    @system("./final_run_narrow.sh");
                            }
                            else
                            {
                                    system("cp /home/vns4483/public_html/final_run_wide.sh /home/vns4483/public_html/uploads/$dirname");
                                    chdir ("/home/vns4483/public_html/uploads/$dirname");
                                    @system("./final_run_wide.sh");
```

```php
                                }
                                #$path1='dcrr.tar.gz';
                                if(file_exists('dcrr.tar.gz'))
                                {

#chdir("/home/vns4483/public_html/$dirname/dcrr/");
                                        if (file_exists('dcrr/output.fig'))
                                        {
                                                echo "You can view your result by clicking the link below.<br>";
        $message="http://tortellini.bioinformatics.rit.edu/vns4483/uploads/$dirname/dcrr.tar.gz";
                                                $header="Click on the link to download the DCRR of the protein structure";
                                                echo "<a href='http://tortellini.bioinformatics.rit.edu/vns4483/uploads/$dirname/dcrr.tar.gz'>Result</a>";
                                                echo "<br><br>";
                                                echo "Result will also be emailed to you at $email<br><br>";
                                                $mail_sent = @mail( $email, "Result from Protein DCRR Web Server", $message, $header);
                                                echo $mail_sent ? "Mail sent" : "Mail failed. Could not send the email.";

                                                #echo "<br>Could not process your request at this time.Check the file for PDB format and try again later<br>";
                                        }
                                        else
                                        {
                                                #echo "You can view your result in the link marked Result<br>";
                                                #echo "You can view your result in the link marked Result<br>";
                                                #echo "<a href='http://tortellini.bioinformatics.rit.edu/vns4483/uploads/$dirname/dcrr.tar.gz'>'Result'</a>";
                                                echo "<br>Could not process your request at this time.Check the file for PDB format and try again later<br>";
                                        }

                                }
                                else
                                {
                                        echo "<br>Check the format of the uploaded file<br>";

                                }
                                #$d="/home/vns4483/public_html/try/$dirname/";
                                #echo("$d");
                                #$command="matlab -sd" .$d. "-nodesktop -nosplash -r colde";
                                #exec($command);
```

```php
                            #echo "<a
href='http://tortellini.bioinformatics.rit.edu/vns4483/try/$dirname/dcrr.tar.
gz'>'Result'</a>";
                            #system("rm -r /home/vns4483/public_html/try");
                            #system("rm -r /home/vns4483/public_html/try/1cns");
                            #system("rm -r /home/vns4483/public_html/try/103m");

                    }
                    else
                    {
                            echo "Error: A problem occurred during file upload!";
                    }
                    }
            else
            {
                     echo "Error: File ".$_FILES["uploadfile"]["name"]."
already exists";
                    }
        }
        else
        {
            echo "Error: Only .txt files  under 1000 Kb are accepted for
upload";
        }// size check ends here

} #topmost else ends here

echo "<br></br>";
echo"Contact Vrunda Sheth <vns4483@rit.edy> Or Vicente Reyes <vmrsbi@rit.edu>
for reporting bugs <br>";
mysql_close();
function input_val($data)
{
    $data = trim($data);
    $data = stripslashes($data);
    $data = htmlspecialchars($data);
    return $data;
}
############################################################################
("digitalpoints",2005)
#######################################################################
"
function check_email($email)
        {
                if(@ereg("^[_a-z0-9-]+(\.[_a-z0-9-]+)*@[a-z0-9-]+(\.[a-z0-9-
]+)*(\.[a-z]{2,4})$", $email))
                {
                     $url = substr(strrchr($email, "@"), 1);
                         if (strstr($url, "/"))
                    {
                        $url = explode("/", $url, 2);
                        $url[1] = "/".$url[1];
                        }
                    else
                    {
                        $url = array($url, "/");
```

```
                                }
                                $fh = @fsockopen($url[0], 80);
                                if ($fh)
                    {
                                @fputs($fh,"GET ".$url[1]." HTTP/1.1\nHost:".$url[0]."\n\n");
                                if (@fread($fh, 22) == "HTTP/1.1 404 Not Found") { return FALSE; }
                                else { return TRUE;     }
                                }
                    else { return FALSE;}
            }
            else
            {
                                return FALSE;
            }
        }
?>
"
```